\documentclass[iop,12pt]{emulateapj}
\usepackage{apjfonts}
\usepackage{natbib}
\usepackage{amsfonts}
\usepackage{amsmath}
\usepackage{multirow}
\usepackage{enumerate}
\usepackage[usenames]{color}

\def\Dwa{$\,$\uppercase\expandafter{\romannumeral5}$\,$}

\def\sless{\lower2pt\hbox{$\buildrel {\scriptstyle <}
   \over {\scriptstyle\sim}$}}

\def\sgreat{\lower2pt\hbox{$\buildrel {\scriptstyle >}
   \over {\scriptstyle\sim}$}}
\def\sharpnull#1{}

\def\sref{\S\ref}
\def\fref{Fig.~\ref}

\def\tref{Table~\ref}
\def\esref{Equations~\ref}
\def\eref{Equation~\ref}

\newcommand{\code}[1]{\texttt{#1}}

\begin{document}
\slugcomment{Accepted for Publication in The Astrophysical Journal, November 16, 2012}

\title{The Progenitor Dependence of the Preexplosion Neutrino Emission\\
in Core-Collapse Supernovae}

\author{Evan O'Connor\altaffilmark{1} and Christian
  D. Ott\altaffilmark{1,2 $\star$}} \altaffiltext{1}{TAPIR, Mailcode 350-17,
  California Institute of Technology, Pasadena, CA 91125,
  evanoc@tapir.caltech.edu, cott@tapir.caltech.edu} 
  \altaffiltext{2}{Kavli IPMU, University of Tokyo, Kashiwa, Japan}
  \altaffiltext{$\star$}{
  Alfred P. Sloan Research Fellow}

\begin{abstract}
We perform spherically-symmetric general-relativistic simulations of
core collapse and the postbounce preexplosion phase in 32 presupernova
stellar models of solar metallicity with zero-age-main-sequence masses
of $12\,M_\odot$ to $120\,M_\odot$.  Using energy-dependent
three-species neutrino transport in the two-moment approximation with
an analytic closure, we show that the emitted neutrino luminosities
and spectra follow very systematic trends that are correlated with the
compactness ($\sim M/R$) of the progenitor star's inner regions via
the accretion rate in the preexplosion phase. We find that these
qualitative trends depend only weakly on the nuclear equation of
state, but quantitative observational statements will require
independent constraints on the equation of state and the rotation rate
of the core as well as a more complete understanding of neutrino
oscillations.  We investigate the simulated response of water
Cherenkov detectors to the electron antineutrino fluxes from our
models and find that the large statistics of a galactic core collapse
event may allow robust conclusions on the inner structure of the
progenitor star.
\end{abstract}


\section{Introduction}
\label{sec:intro}

The radial instability of the electron-degenerate, Chandrasekhar-mass
core marks the beginning of the final episode in the life of a massive
star with zero-age main-sequence (ZAMS) mass in the range $\sim8\,
M_\odot - 130\, M_\odot$. Collapse ensues and, once fully dynamical,
separates the core into a subsonically homologously contracting inner
core and a supersonically collapsing outer core.  At nuclear density,
the repulsive component of the nuclear force leads to a stiffening of
the equation of state (EOS). This stabilizes the inner core, which
overshoots its new equilibrium, then rebounds into the outer core,
launching a strong hydrodynamic shock wave from its edge.  This
instant in time is referred to as \emph{core bounce}. The inner core
has a mass of $\sim0.5\,M_\odot$ at bounce and this material becomes
the unshocked core of the protoneutron star. The shock formed at core
bounce propagates into the outer core, but the dissociation of nuclei
in the accreted material into neutrons and protons and the electron
capture on free protons in the region behind the shock (the
\emph{postshock} region) soon sap its might, driving it into
submission to the ram pressure of accretion. The shock stalls and
turns into a standing accretion shock that must be revived to unbind
the stellar envelope and drive a core-collapse supernova explosion.

Neutrinos play a pivotal and dominant role in stellar collapse and
core-collapse supernovae. Neutrinos and antineutrinos of all flavors
carry away the $\sim$$300\,\mathrm{B}$ ($= 3\times
10^{53}\,\mathrm{ergs}$) of gravitational binding energy of the
remnant neutron star over tens of seconds after core bounce. Aided by
multi-dimensional fluid instabilities, they probably deposit, within a
few hundred milliseconds after bounce, sufficient energy in the region
behind the stalled shock to revive the shock and make
typical $\sim$$1\,\mathrm{B}$ core-collapse supernova explosion
(\citealt{janka:07,mueller:12a} and references therein).
Only hyper-energetic (i.e., $\mathcal{O}(10)\,\,\mathrm{B}$)
explosions may require a different mechanism \citep{ugliano:12}, e.g.,
rapid rotation combined with strong magnetic fields, which may lead to
energetic jet-driven explosions (e.g., \citealt{burrows:07b}).

For a galactic or near-extragalactic core-collapse supernova,
neutrinos offer the unique possibility of directly observing the
dynamics and thermodynamic conditions prevalent in the supernova core.
Together with gravitational waves (see, e.g.,
\citealt{ott:09,kotake:11b}) they will herald the next nearby
supernova possibly hours before any telescope sensitive to
electromagnetic waves will notice the event. In the probable case that
the next galactic supernova occurs in a dust-enshrouded region and/or
close or behind the galactic center, the supernova may be impossible
to observe in broad bands of the electromagnetic spectrum, making
neutrino and gravitational-wave observations even more important.

The observation of neutrinos from SN 1987A in the Large Magellanic
Cloud \citep{hirata:87,bionta:87,alekseev:87} confirmed the basic
picture of core collapse and early protoneutron star evolution (e.g.,
\citealt{sato:87, bruenn:87, burrows:86, burrows:87, burrows:88, arnett:89,
  jegerlehner:96, loredo:02, yueksel:07, pagliaroli:09b} and
references therein), but the small number and poor timing of the
observed interactions did not allow far-reaching and robust conclusions on
core-collapse supernova dynamics and the involved neutrino physics,
nuclear physics and astrophysics.

The situation will be completely different when the neutrino burst
from a galactic core-collapse supernova reaches current and
near-future neutrino detectors on Earth.  Super-Kamiokande
\citep{fukuda:03,ikeda:07}, IceCube \citep{icecube:11sn}, LVD
\citep{aglietta:92,lvd:11}, Borexino \citep{borexino:09,cadonati:02}, KamLAND 
\citep{piepke:01kamland}, SNO+ \citep{kraus:10snoplus}, No$\nu$a
\citep{davies:11nova} and others will together see many thousands of
neutrinos from a core collapse event at $10\,\mathrm{kpc}$
\citep{scholberg:12}.  Distance estimates based on the observed
neutrino flux \citep{kachelriess:05}, sky localization
\citep{beacom:98,tomas:03}, and triggering of gravitational-wave
searches by the reconstruction of the time of core bounce
\citep{pagliaroli:09,halzen:09} will likely all be possible.  

Well-timed high-statistics coincident neutrino observations will allow
to probe in detail a broad range of supernova astrophysics, nuclear
physics, and neutrino physics (see
\citealt{burrows:92b,wurm:12,raffelt:10ov} for overviews). Fast
characteristic temporal variations in the preexplosion neutrino fluxes
would be tell-tale signs of multi-dimensional fluid instabilities in
the postshock region \citep{ott:08,marek:09b,lund:10,brandt:11} and/or
early postbounce ring-down oscillations of a rapidly spinning
protoneutron star \citep{ott:12a}. While artificially-driven,
spherically-symmetric core collapse simulations suggest that a sudden
deep drop of the accretion-driven component of the neutrino luminosity
(primarily in $\nu_e$ and $\bar{\nu}_e$) within a few hundred
milliseconds would indicate the onset of explosion (e.g.,
\citealt{burrows:92b,fischer:12a,fischer:10}), the situation is likely
different in nature as multidimensional simulations show a much
shallower drop due to simultaneous accretion and explosion of
material.  This leads to a persistent accretion luminosity even after
the explosion has begun \citep{mueller:12a,mueller:12b}. The spectral
characteristics and long-term spectral evolution of the neutrino flux
could provide important constraints on the nuclear EOS
\citep{roberts:12,marek:09b} and/or the spin of the progenitor core
\citep{ott:08,marek:09}. The ability of some detectors to distinguish
interactions of different neutrino flavors would lead to constraints
on the neutron-to-proton ratio in the neutrino-driven wind phase,
allowing an observational test of core-collapse supernovae as
potential sites for $r$-process nucleosynthesis
\citep{huedepohl:10,fischer:10,wurm:12}.

The neutrino signature of core collapse, of the subsequent
core-collapse supernova evolution and of the protoneutron star cooling
phase, is invariably intertwined with neutrino oscillation
physics. The robustness of all of the above mentioned observational
conclusions will depend on our understanding of the impact of neutrino
flavor oscillations. Neutrinos propagating from their emission site to
detectors on Earth may experience (\emph{i}) so-called vacuum
oscillations driven by neutrino mass differences
\citep{Pontecorvo:1967fh}, (\emph{ii}) oscillations mediated by a
resonance in $\nu$--$e^-$ scattering (the Mikheyev-Smirnov-Wolfenstein
[MSW] effect; \citealt{Mikheev:86,wolfenstein:78}), and
(\emph{iii}) oscillations due to $\nu$--$\nu$ scattering
(\citealt{Pantaleone:1992eq}; see \citealt{duan:10} for a
review). Vacuum and MSW oscillations are well understood and their
outcomes depend essentially only on neutrino mixing parameters, in
particular the neutrino mass hierarchy and the mixing angles. The
$\nu$--$\nu$-scattering driven oscillations, on the other hand, have a
non-linear Hamiltonian that may lead to so-called collective
oscillations with very complex spatial and temporal outcome that
remains to be fully understood (see, e.g.,
\citealt{hannestad:06,duan:07a,fogli:09,Dasgupta:2007ws,duan:10} and
references therein). However, a number of recent studies suggest that
collective oscillations may be completely or at least partially
suppressed in the preexplosion accretion phase of ordinary
core-collapse supernovae \citep{chakraborty:11prl,
  Chakraborty:2011gd, sarikas:12}, but see \cite{cherry:12} and
\cite{dasgupta:12a} for discrepant results.

Provided that collective oscillations can be ignored in the 
preexplosion phase and that the $\theta_{13}$ mixing angle
indeed has the large value suggested by recent measurements
\citep{dayabay:12}, the neutrino mass hierarchy may be inferred from
the qualitative shape of the early postbounce neutrino signal
 \citep{kachelriess:05,serpico:12}. 

A preexplosion accretion phase with suppressed collective oscillations
would also offer the opportunity to probe the structure of the
progenitor star on the basis of the observed neutrino signal.  The
details of the preexplosion neutrino emission have been discussed
carefully, e.g., by \cite{thompson:03} and
\cite{liebendoerfer:04} and we shall not repeat them here. It is,
however, necessary to outline its most salient 
features. For simplicity, we neglect neutrino oscillations in the
following.

In core collapse and in the subsequent postbounce evolution, emission
of $\nu_e$ and $\bar{\nu}_e$ occurs via charged and neutral currents,
while heavy-lepton neutrinos $\nu_x = \{\nu_\mu, \bar{\nu}_\mu,
\nu_\tau, \bar{\nu}_\tau\}$ are created exclusively via thermal
neutral-current pair processes. Before core bounce, only $\nu_e$ are
emitted from electron capture in the collapsing core. Milliseconds
after core bounce, the shock breaks out of the $\nu_e$ neutrinosphere
(where the optical depth is $\tau_{\nu_e} \approx 2/3$) and a strong
burst of $\nu_e$ is emitted for $\sim$$20\,\mathrm{ms}$ from rapid
electron capture on the freshly abundant free protons behind the
shock.  $\nu_x$ are copiously created in the hot interior of the
protoneutron star after bounce and begin to diffuse out, leading to a
steep rise, quick leveling and subsequent slow decay of the $\nu_x$
luminosity ($L_{\nu_x}$). $\bar{\nu}_e$ production via charged-current
positron capture is initially suppressed due to the high degeneracy of
the electrons. The latter is partially lifted after bounce at the
moderate-density, hot edge of the protoneutron star and
$L_{\bar{\nu}_e}$ rises, reaching or surpassing the value at which
$L_{\nu_e}$ levels off after the neutronization burst decays. The
subsequent preexplosion luminosity can roughly be split into a
diffusive component from the core and accretion luminosity ($\propto G
M[R_\nu] \dot{M} / R_\nu$, where $R_\nu$ is an approximate
neutrinosphere radius) from or from above the neutrinosphere
\citep{burrows:88}. $L_{\nu_x}$ is primarily diffusive, while
$L_{\nu_e}$ and $L_{\bar{\nu}_e}$ are dominated by accretion.  In
general, $L_{\nu_e} \approx L_{\bar{\nu}_e} > L_{\nu_x}$, but $4
L_{\nu_x} = L_{\nu_\mu} + L_{\bar{\nu}_\mu} + L_{\nu_\tau} +
L_{\bar{\nu}_\tau} > L_{\nu_e} + L_{\bar{\nu}_e}$. $\nu_x$ have the
lowest opacity, since they interact only via neutral currents. They
decouple from matter at the smallest radii and highest temperatures
and thus have the highest average energies
$\langle\epsilon_\nu\rangle$. $\bar{\nu}_e$ have a slightly lower
opacity than $\nu_e$, leading to the well established neutrino energy
hierarchy in the preexplosion phase $\langle\epsilon_{\nu_x}\rangle >
\langle\epsilon_{\bar{\nu}_e}\rangle >
\langle\epsilon_{\nu_e}\rangle$. The mean energy of all species grows
with increasing postbounce time, reflecting the recession of the
neutrinospheres due to the contraction of the protoneutron star.

Considering that the accretion luminosity will scale with the
postbounce accretion rate $\dot{M}$, one would naturally expect an
increase of the detected neutrino interactions with increasing mass of
the stellar core. Since higher accretion rates correspond to more
material compressing and settling more rapidly on the protoneutron
star, the latter's outer regions will be hotter. Thus the thermal
neutral-current emission will be enhanced, leading to higher
luminosities and higher mean neutrino energies.

The variation of the preexplosion neutrino signal with progenitor star
ZAMS mass was first discussed by \cite{woosley:86} based on the
pioneering simulations of \cite{wilson:85} and \cite{wilson:86}. These
authors provided total emission characteristics and spectra that show
a systematic increase of total energy emitted in neutrinos and mean
$\bar{\nu}_e$ energy with ZAMS mass in the range from $10 -
25\,M_\odot$.  \cite{mayle:87}, before SN 1987A, carried out
simulations of a range of progenitor stars with ZAMS mass in the range
$12-100\,M_\odot$. They found that the $\nu_e$ neutronization burst
shows little dependence on the progenitor, due to the rather universal
homologous collapse and bounce dynamics. Furthermore, they mentioned,
though did not discuss in detail, that the luminosities and mean
neutrino energies increase as a function of iron core mass (and not
ZAMS mass). A more detailed and clear physical discussion was provided
by \cite{bruenn:87}, who contrasted the predicted neutrino signal from
the early postbounce phase in two different progenitor core models
with neutrino observations of SN 1987A. He noted that there are
significant uncertainties in connecting a given ZAMS mass to
precollapse structure. Instead of a progenitor with an associated ZAMS
mass, he considered a massive (and high-entropy) $2.05$-$M_\odot$ iron
core model and a lower-mass (and lower-entropy) $1.35$-$M_\odot$ iron
core model in his spherically-symmetric (1D) neutrino
radiation-hydrodynamic simulations. He showed that the more massive
core leads to a consistently higher $\bar{\nu}_e$ luminosity in both
the accretion and diffusion sectors. The water Cherenkov detectors
that observed neutrinos from SN 1987A are most sensitive to the
inverse beta decay (IBD) reaction $\bar{\nu}_e + p \rightarrow n +
e^+$. \cite{bruenn:87} predicted a factor of two difference in the
integrated number of early IBD interactions between the massive and
the low-mass core in these detectors.  He concluded that the neutrino
signal observed by these detectors from SN 1987A was most consistent
with the low-mass core. \cite{burrows:88}, who carried out a parameter
study of quasi-hydrostatic protoneutron star cooling, considering
various initial masses, ad-hoc accretion rates, and different nuclear
EOS, found a similar trend.  He showed that more massive cores, higher
accretion rates, and softer EOS lead to stronger, higher-energy
neutrino emission. Some of his strongest emitters were cases in which
eventually a black hole was formed.

Liebend\"orfer and collaborators carried out a sequence of studies of
the progenitor dependence of the neutrino signal using modern
general relativistic 1D radiation-hydrodynamics simulations
\citep{liebendoerfer:01c,liebendoerfer:02b,liebendoerfer:03,
  liebendoerfer:04}. They showed that the $\nu_e$ neutronization burst
is indeed almost independent of progenitor structure (as first
suggested by \citealt{mayle:87}). They also qualitatively and
quantitatively connected the evolution of the postbounce preexplosion
luminosity to the postbounce accretion rate, but did not discuss
observational implications. Their results were corroborated by
similarly sophisticated subsequent studies of
\cite{kachelriess:05,buras:06b,fischer:09a,sumiyoshi:08,fischer:10,fischer:12a,
  serpico:12}. Of these, \cite{buras:06b} presented the most
comprehensive analysis and also compared between 1D and axisymmetric
(2D) results. They found that in 2D, convection in the protoneutron
star alters the structure of the latter, affecting the neutrino
emission starting $\sim$$100\,\mathrm{ms}$ after bounce, but
preserving the overall systematics with accretion
rate. \cite{buras:06b} also were the only authors to suggest that the
accretion-rate dependence of luminosity and total emitted energy in
the preexplosion phase could be used to infer the structure of the
progenitor. The other studies, being focused on aspects such as
neutrino oscillations, black hole formation, or the late-time
post-explosion evolution, did not consider observational consequences.

\cite{thompson:03}, using a limited set of three progenitor models
(\{$11$, $15$, $20$\}$\, M_\odot$ at ZAMS), found similar systematics
as the aforementioned studies, but also carried out an analysis of the
expected signal in various neutrino detectors in the first
$250\,\mathrm{ms}$ after bounce.  They computed IBD interaction rates
for their $20$-$M_\odot$ and $11$-$M_\odot$ and found a factor of two
more IBD interactions for the former, which would allow a high-confidence
distinction between these progenitors for a galactic core collapse
event.  However, their $15$-$M_\odot$ model yielded a postbounce
neutrino signal very similar to that of their $11$-$M_\odot$ model and
would be indistinguishable by neutrino observations alone. This
suggests that ZAMS mass is not a good parameter to describe
presupernova stellar structure (cf., \citealt{bruenn:87}).

In this article, we present a fresh look at the progenitor dependence
of the neutrino signature in the preexplosion accretion phase of
core-collapse supernovae. We perform 1D general relativistic
radiation-hydrodynamics core collapse simulations of 32 progenitor
models from the single-star solar-metallicity presupernova model suite
of \cite{woosley:07} and follow the postbounce preexplosion evolution
for $450\,\mathrm{ms}$.  In ZAMS mass, these models range from
$12\,M_\odot$ to $120\,M_\odot$, but guided by the previous results
discussed in the above, we choose not to parameterize our simulations
by ZAMS mass. Instead we employ the compactness parameter $\xi_M \sim
M/R(M)$ (for a relevant mass scale $M$, measured at the time of
bounce). As shown in \cite{oconnor:11}, and further explored in
\cite{ugliano:12}, $\xi_M$ is a quantitative stellar structure
parameter that describes the postbounce accretion evolution to a
remnant mass scale $M$. We demonstrate that the preexplosion neutrino
emission is very well parameterized by the compactness. The
preexplosion luminosities and mean energies of all neutrino species
increase essentially monotonically with increasing $\xi_M$. We compute
predicted integrated IBD interactions for a galactic core-collapse
supernova in the Super-Kamiokande detector and show that the clear
systematics governed by $\xi_M$ carries over to observation, even when
standard MSW neutrino oscillations are taken into account. Our results
thus indicate that -- in the absence of complicated collective
neutrino oscillations -- a high-statistics detection of neutrinos from
the preexplosion phase will allow, in principle, a tight constraint of
the compactness of the progenitor star's core.  This, however, will
require knowledge of the nuclear EOS and of the rotation rate of the
collapsed core, since, as we show, both can dilute the otherwise clear
compactness-dependent neutrino emission systematics.

This paper is structured as follows. In \S\ref{sec:methods}, we
discuss our general relativistic hydrodynamics code {\tt GR1D} and
introduce its extension to neutrino radiation-hydrodynamics in the
two-moment approximation, {\tt nuGR1D}. The initial models and the
employed EOS are discussed in \S\ref{sec:models}. In
\S\ref{sec:boltzmanncompare}, we present results from a benchmark
collapse and postbounce simulation that allows us to compare with the
previously published code comparison of \cite{liebendoerfer:05} to
assess {\tt nuGR1D}'s ability to reproduce results of full Boltzmann
neutrino transport. We present the results of our simulations in
\S\ref{sec:results}, analyze the dependence of the neutrino signal on
progenitor compactness, discuss predicted IBD signals from a galactic
core collapse event in the Super-Kamiokande detector, and explore
potential degeneracies introduced by EOS and rotation. Finally, in
\S\ref{sec:discussion}, we critically summarize our work and conclude
by contrasting our results with the early neutrino signal observed
from SN 1987A.

\section{Methods}
\label{sec:methods}

We make use of the open-source 1D general relativistic hydrodynamics
code \code{GR1D} (\citealt{oconnor:10}; available at {\tt
  http://www.stellarcollapse.org}) outfitted with an energy-dependent
multi-species M1 neutrino transport scheme in which the zeroth and
first moments of the neutrino distribution function are evolved. We
refer the reader to \cite{oconnor:10} for details on \code{GR1D} and
describe in the following our current implementation of the transport
scheme. For this first application, we neglect the
computationally-expensive energy-coupling neutrino interactions and
transport terms -- these terms are undoubtedly important for making
highly accurate predictions of the neutrino signature (see, e.g.,
\citealt{lentz:12b,lentz:12a}), but are unlikely to affect the general
trends we observe. We will address them in future work, but provide a
discussion on the consequences of neglecting these terms via a
comparison to full Boltzmann neutrino transport simulations in
\sref{sec:boltzmanncompare}.

Our M1 scheme closely follows \cite{shibata:11}, who formulate the M1
evolution equations in a closed covariant form. The scheme is
simplified greatly by neglecting the energy-coupling terms.  This
further requires that the velocity dependent terms are also
ignored. In this limit, and using the Schwarzschild-like metric and
radial-polar slicing of \code{GR1D} and setting $G=c=M_\odot = 1$, the
coordinate frame evolution equations for the neutrino energy density,
$E_{(\nu)}$, and the neutrino flux vector, $F_{r,(\nu)}$, simplify
from Equations~3.37 and 3.38 of \cite{shibata:11} to
\begin{equation}
\partial_t E_{(\nu)} + \frac{1}{r^2}\partial_r\left(\frac{\alpha
  r^2}{X^2}F_{r,(\nu)}\right) = \alpha^2
\mathcal{S}^t_{(\nu)}\,,\label{eq:energyevolution}
\end{equation}
and
\begin{equation}
\partial_t F_{r,(\nu)} + \frac{1}{r^2}\partial_r\left(\frac{\alpha
  r^2}{X^2}P_{rr,(\nu)}\right) = \alpha X^2 \mathcal{S}^r_{(\nu)} + \alpha
\frac{E_{(\nu)}(1-p_{(\nu)})}{r}\,,\label{eq:fluxevolution}
\end{equation}
where $\mathcal{S}^\alpha$ is the neutrino interaction source term
(see below), $\alpha$ is the lapse function and $X = (1 - 2
M(r)/r)^{-1/2}$.  $P_{rr,(\nu)}$ is the neutrino pressure tensor and
is taken to be an interpolation between the two limiting cases of free
streaming and diffusion.  We follow \cite{shibata:11}, who express
$P_{ii,(\nu)}$ as
\begin{equation}
P_{ii,(\nu)} = \frac{3p_{(\nu)}-1}{2}P_{ii,(\nu),\mathrm{thin}} +
\frac{3(1-p_{(\nu)})}{2}P_{ii,(\nu),\mathrm{thick}}\,,
\end{equation}
where $p_{(\nu)}$ is the Eddington factor, taken here to be the maximum
entropy closure in a closed, analytic form
\citep{minerbo:78,cernohorsky:94},
\begin{equation}
p_{(\nu)} = \frac{1}{3} +
\frac{f_{(\nu)}^2}{15}(6-2f_{(\nu)}+6f_{(\nu)}^2)\,\,.
\end{equation}
In the no-velocity limit for \code{GR1D},
$f_{(\nu)}=|F_{r,(\nu)}/(E_{(\nu)}X)|$. The free streaming and
diffusion limits of the neutrino pressure tensor are
$P_{rr,(\nu),\mathrm{thin}} = E_{(\nu)} X^2$ and
$P_{ii,(\nu),\mathrm{thick}} = g_{ii}E_{(\nu)}/3$, respectively.

We set out to solve the system of equations via standard hyperbolic
methods borrowed from conservative hydrodynamic schemes
\citep{pons:00}. Transport variables live at cell centers and we
employ piece-wise linear reconstruction to cell interfaces with van
Leer's limiter \citep{vanleer:77} and the HLLE approximate Riemann
solver \citep{HLLE:88} for calculating the intercell fluxes.  A
complication arises when solving the neutrino moment equations in the
high-opacity limit.  In this case, the standard fluxes returned from
the Riemann solver are dominated by a numerical diffusion term.  We
follow \cite{audit:02} and modify the fluxes to correct for this. 
Essentially, the modified fluxes return the diffusion-limit flux in
the high opacity limit. The modified form of the neutrino energy
density flux through the $i+1/2$ interface is given by
\begin{equation}
F_{r,(\nu)}^{i+1/2} = \frac{\tilde a^+F_{r,(\nu)}^{i,R} - \tilde
  a^-F_{r,(\nu)}^{i+1,L} + \epsilon_{(\nu)}\tilde a^+ \tilde a^-
  (E_{(\nu)}^{i+1,L} - E_{(\nu)}^{i,R})}{\tilde a^+ - \tilde a^-}\,.
\end{equation}
The corresponding modified interface flux for the neutrino flux
evolution equation is,
\begin{equation}
P_{rr,(\nu)}^{i+1/2} = \epsilon_{(\nu)}\tilde P_{rr,(\nu)}^{i+1/2} +
(1-\epsilon_{(\nu)}^2)(P_{(\nu)}^{i+1,L} + P_{(\nu)}^{i,R})/2\,,
\end{equation}
with
\begin{equation}
\tilde P_{rr,(\nu)}^{i+1/2} = \frac{\epsilon_{(\nu)}(\tilde
  a^+P_{rr,(\nu)}^{i,R} - \tilde a^-P_{rr,(\nu)}^{i+1,L}) + \tilde a^+
  \tilde a^-(F_{(\nu)}^{i+1,L} - F_{(\nu)}^{i,R})}{\tilde a^+ - \tilde
  a^-}\,.
\end{equation}

In these equations, $\epsilon_{(\nu)}$ controls the modification to
the fluxes to account for the high opacity. Following \cite{audit:02},
we take
\begin{equation}
\epsilon_{(\nu)} = \mathrm{min}\left(1,\frac{1}{\kappa_{(\nu)} \Delta
  r}\right)\,\,,
\end{equation}
where $\kappa_{(\nu)}$ is the sum of the scattering and absorptive
opacities.  These opacities are strong functions of energy and are
also species dependent. We note that when $\epsilon_{(\nu)}$ is 1, the
intercell fluxes reduce to the standard HLLE approximation. The
characteristic speeds needed for the HLLE scheme are calculated in the
same spirit as the neutrino pressure tensor
\citep{shibata:11,kuroda:12},
\begin{equation}
\lambda_{(\nu)} = \frac{3p_{(\nu)}-1}{2}\lambda_{(\nu),\mathrm{thin}} +
\frac{3(1-p_{(\nu)})}{2}\lambda_{(\nu),\mathrm{thick}}\,,
\end{equation}
where in the zero velocity limit, $\lambda_{(\nu),\mathrm{thin}} = \pm
\alpha/X$ and $\lambda_{(\nu),\mathrm{thick}} = \pm
\alpha/(\sqrt{3}X)$.  $\tilde a^+$ and $\tilde a^-$ are the maximum
and minimum values, respectively, of these characteristic speeds
evaluated from both the right and left reconstructed
variables.

Finally, the source terms in \esref{eq:energyevolution} and
\ref{eq:fluxevolution} are taken from \cite{shibata:11}.  In the zero
velocity limit,
\begin{eqnarray}
\mathcal{S}^t &=& (\eta_{(\nu)} - \kappa_{a,(\nu)} E_{(\nu)})/\alpha\,,\label{eq:st} \\
\mathcal{S}^r &=& -(\kappa_{a,(\nu)}+\kappa_{s,(\nu)}) F_{r,(\nu)}/X^2\,,
\end{eqnarray}
where $\eta_{(\nu)}$, $\kappa_{a,(\nu)}$, and $\kappa_{s,(\nu)}$ are
the neutrino emissivity, neutrino absorption opacity, and the neutrino
scattering opacity, respectively. We precompute the neutrino
interaction terms for each neutrinos species (we treat $\nu_e$,
$\bar{\nu}_e$ and $\nu_x =
\{\nu_\mu,\bar{\nu}_\mu,\nu_\tau,\bar{\nu}_\tau\}$) and neutrino
energy group in dense tabular form as a function of density $\rho$,
temperature $T$, and electron fraction $Y_e$. We then use linear
interpolation for efficient on-the-fly interpolation.  We include all
standard iso-energetic scattering processes, charged-current
absorption and emission, and thermal pair-production processes
\citep{brt:06,bruenn:85} in the calculation of the neutrino
interaction terms. Since the neutrino--matter interactions for
heavy-lepton neutrinos and antineutrinos are slightly different,
\code{NuLib} averages the two values of the emissivities and
opacities. Our library of neutrino interaction routines, which we call
\code{NuLib}, is open source and available as a GitHub repository at
     {\url http://www.nulib.org}. \code{NuLib} requires an EOS for the
     evaluation of the emissivities and opacities.  Our treatment of
     thermal pair-processes in GR1D warrants some comments.  Since we
     do not currently consider energy (or species) coupling for
     thermal emission processes such as electron--positron
     annihilation to a neutrino--antineutrino pair, we compute an
     emissivity based on the thermal content of the matter ignoring
     any final state neutrino blocking.  To limit the neutrino energy
     density to the equilibrium value (where neutrino-antineutrino
     annihilation rates are in equilibrium with the thermal pair
     production rates), we use Kirchhoff's law to derive an
     \emph{effective} absorption opacity for neutrino--antineutrino
     annihilation from the thermal emissivity,
\begin{equation}
\kappa^\mathrm{eff, thermal}_{a,(\nu)} = \eta^\mathrm{thermal}_{(\nu)} / B_{(\nu)}\,,
\end{equation}
where $B_{(\nu)} = c E_{(\nu)}^3/(2\pi \hbar c)^3
f_{(\nu)}^\mathrm{eq}$ is the thermal energy density of neutrinos with
energy $E_{(\nu)}$ and $f_{(\nu)}^\mathrm{eq}=1/(\exp{[(E_{(\nu)} -
    \mu)/T]}+1)$ is the equilibrium neutrino distribution
function with chemical potential $\mu$. As we shall see, this
method performs well at predicting the thermal neutrino flux of the
heavy-lepton neutrinos during the preexplosion phase.

In \code{nuGR1D}, we first update the hydrodynamic variables to the
$n+1$-th timestep. We then compute the neutrino opacities and
emissivities associated with the updated hydrodynamic variables. We
update the radiation field operator-split. The flux term is solved
explicitly, using the radiation moments of the $n$-th timestep. We
calculate the neutrino--matter interaction terms using the $n+1$
radiation moments via a local implicit update. With the $n+1$
radiation energy density source term, we then update the energy
density and electron fraction of the matter. We use 24 energy groups,
with lowest-energy group centers at 0.5\,MeV and 1.5\,MeV, and then
spaced logarithmically up to 200\,MeV for $\nu_e$, $\bar{\nu}_e$ and
$\nu_x$.  We note that for the highest energy bins it occasionally
occurs that the evolved neutrino flux vector exceeds the evolved
neutrino energy density. This tends to occur in the most dynamic
phases of our simulations and where the opacities vary significantly
from one zone to the next.  When this is the case we limit the
neutrino flux to the neutrino energy density. We extract the radiation
quantities in the coordinate frame at a radius of 500\,km.

\section{Initial Models and Equations of State}
\label{sec:models}

\begin{figure}[t]
\centering
\includegraphics[width=0.97\columnwidth]{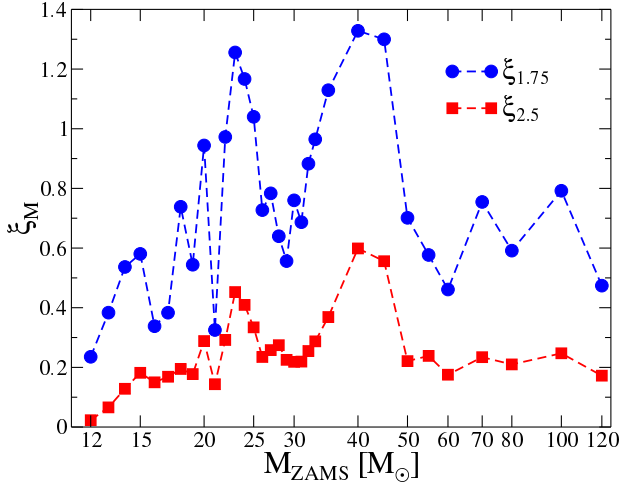}
\caption{Compactness parameters for the 32 considered presupernova
  models of \cite{woosley:07} versus ZAMS mass as evaluated from
  collapse simulations with the LS220 EOS.  We show both $\xi_{1.75}$
  and $\xi_{2.5}$.  The mapping between ZAMS mass and precollapse
  structure is highly non-monotonic, making the former an ill-suited
  parameter for describing progenitor structure in core collapse
  simulations.} \label{fig:xis}
\end{figure}

We employ the most recent non-rotating solar-metallicity single-star
model set from the stellar evolution code \code{KEPLER}
\citep{woosley:07}.  This model set contains the presupernova
configuration of 32 stars ranging in ZAMS mass from 12\,$M_\odot$ to
120\,$M_\odot$. We denote individual models by $sXX$WH07, where $XX$
corresponds to the integer ZAMS mass of the model, e.g., $s12$WH07 is
the $12$-$M_\odot$ model of this model set. In \cite{oconnor:11}, we
investigated this and other model sets in the context of black hole
formation. Under the assumption of a failed core-collapse supernova,
we found a strong empirical relation between the properties of the
presupernova structure and the evolution of the failing supernova,
e.g., the time to black hole formation.  This led to a clear
prediction: If we observe black hole formation in a failed
core-collapse supernova via neutrinos, the lifetime of the
protoneutron star (and thus of the neutrino signal) relays direct
information about the presupernova structure.  However, such a
prediction, (\emph{i}) requires a failed supernova, which may not be
the norm, and (\emph{ii}) has a strong dependence on the nuclear
equation of state.  The empirical parameter introduced in
\cite{oconnor:11} is the compactness of the progenitor, measured at
the time of core bounce.  It is an inverse measure of the radial
extent of a given mass coordinate at the time of bounce,
\begin{equation}
  \xi_{M} = {M / \,M_\odot  \over R(M_\mathrm{bary} =
    M) / 1000\,\mathrm{km}}\Big|_{t =t_{\mathrm{bounce}}}\,,\label{eq:bouncecompactness}
\end{equation}
where $R(M_\mathrm{bary}=M)$ is the radial coordinate that encloses a
baryonic mass of $M$ at the time of core bounce. In \cite{oconnor:11},
we chose $M = 2.5\,M_\odot$, since this is the relevant mass scale for
black hole formation, i.e., a typical maximum baryonic mass at which a
range of EOS can no longer support a neutron star against gravity.  In
this study, we primarily use $\xi_{1.75}$. The motivation for this is
that during the postbounce preexplosion phase, the relevant mass
scale, especially for models with relatively small compactness, is
much less than 2.5\,$M_\odot$. In this study, we choose
1.75\,$M_\odot$ because this is close to the average baryonic mass
inside the shock at 200--300\,ms after bounce for all models: in the
two extreme models that span the space in compactness parameter (model
s12WH07, [$\xi_{1.75} = 0.24$ and $\xi_{2.5} = 0.022$], on the lower
end; model s40WH07 [$\xi_{1.75} = 1.33$ and $\xi_{2.5} = 0.59$] on the
upper end), the baryonic mass accreted through the shock at 250 ms
after bounce is 1.45 M and 2.05 M, respectively. We further justify
our motivation of using $\xi_{1.75}$ over $\xi_{2.5}$ in
\sref{sec:trends}.  In \fref{fig:xis}, we plot both $\xi_{1.75}$ and
$\xi_{2.5}$ versus ZAMS mass for all 32 considered
models. $\xi_{1.75}$ is provided in \tref{tab:results} for all models.

For \fref{fig:xis}, one notes that while $\xi_{1.75}$ and $\xi_{2.5}$
differ quantitatively, there is no significant qualitative difference
between them.  The overall trends transcending individual models
remain, including the two regions of high compactness near
  22--25\,$M_\odot$ and 35--45\,$M_\odot$. $\xi_{1.75}$ simply
provides a more fine-grained parameterization at the lower mass scale
relevant in the first few hundred milliseconds after bounce. Note,
however, that there are a few  models that have similar
$\xi_{2.5}$, but rather different density structure at small enclosed
masses and radii and, hence, a different $\xi_{1.75}$. Models
$s$14WH07 and $s$16WH07 are examples.

In this study we perform core collapse simulations with each
progenitor and two EOS. We use the EOS of \cite{lseos:91} with a
nuclear incompressibility of 220\,MeV. The LS220 EOS is based on a
compressible liquid-drop model of the nucleus.  Of the publicly
available nuclear EOS, the LS220 EOS best matches the constraints from
nuclear theory and astrophysical observations (see Fig.~1 of
\citealt{ott:12hanse} and
\citealt{demorest:10,hebeler:10,steiner:10,oezel:10b}).  We also
employ the relativistic mean field EOS of \cite{hshen:11} that is
based on the TM1 parameter set.  It is very different from the LS220
EOS.  The maximum neutrino-less $\beta$-equilibrium cold neutron star
gravitational masses are 2.04\,$M_\odot$ and 2.24\,$M_\odot$ for the
LS220 and HShen EOS, respectively.  The radius of a neutrino-less
$\beta$-equilibrium cold neutron star with a gravitational mass of
1.4\,$M_\odot$ using the LS220 EOS is 12.7\,km. For the HShen EOS, the
corresponding radius is 14.6\,km.  For details on our
particular implementation and the treatment of the low-density EOS, we
refer the reader to \cite{oconnor:10,oconnor:11}.  The EOS tables,
reader and interpolation routines are available from {\tt
  http://www.stellarcollapse.org}.

\section{Comparison of \code{nuGR1D} to Boltzmann transport}
\label{sec:boltzmanncompare}

Since our implementation of neutrino transport is new and
approximate, a comparison with published results of full Boltzmann
neutrino transport is warranted. This will allow us to test the
ability of our code to reproduce the neutrino luminosities and
spectral properties in the preexplosion phase.

We compare \code{nuGR1D} with the results of \cite{liebendoerfer:05},
a comparison study between two Boltzmann neutrino transport
codes\footnote{The numerical data from this study are available online
  at
  \url{\mbox{http://iopscience.iop.org/0004-637X/620/2/840/fulltext/datafiles.tar.gz}}}.
The two codes, \mbox{{\it Agile}-\code{BOLTZTRAN}}
\citep{liebendoerfer:04}, and \code{VERTEX} \citep{rampp:02}, approach
the neutrino transport problem in very different ways. Their results
compare very well in the Newtonian limit, but show significant
quantitative differences in the general relativistic case. Subsequent
modifications to the approximate general relativistic potential used
in \code{VERTEX} \citep{marek:05} have since removed many of the
quantitative differences between the codes.  The general relativistic
test case of \cite{liebendoerfer:05} was the collapse and early
postbounce evolution of a $15$-$M_\odot$ (at ZAMS) solar-metallicity
progenitor of \cite{ww:95}, referred to as model $s$15WW95 in the
following.  They employed the LS180 EOS \citep{lseos:91} and a
baseline set of neutrino-matter interactions, including coupling of
energy groups via inelastic scattering processes
\citep{bruenn:85,liebendoerfer:05}.  We repeat their test here, using 
the same initial conditions and EOS, with
our current approximations and compare the neutrino observables.  We
emphasize again that the current version of \code{nuGR1D} lacks
inelastic neutrino--electron scattering and velocity dependent
transport terms. Both are included in the simulations of
\cite{liebendoerfer:05}.  Our transport scheme evolves only the zeroth
and first moment of the neutrino distribution function, using an
analytic closure to truncate the series of moment equations, whereas
\cite{liebendoerfer:05} solve the full Boltzmann equation for neutrino
transport.

\begin{figure}[t]
\centering
\includegraphics[width=1.0\columnwidth]{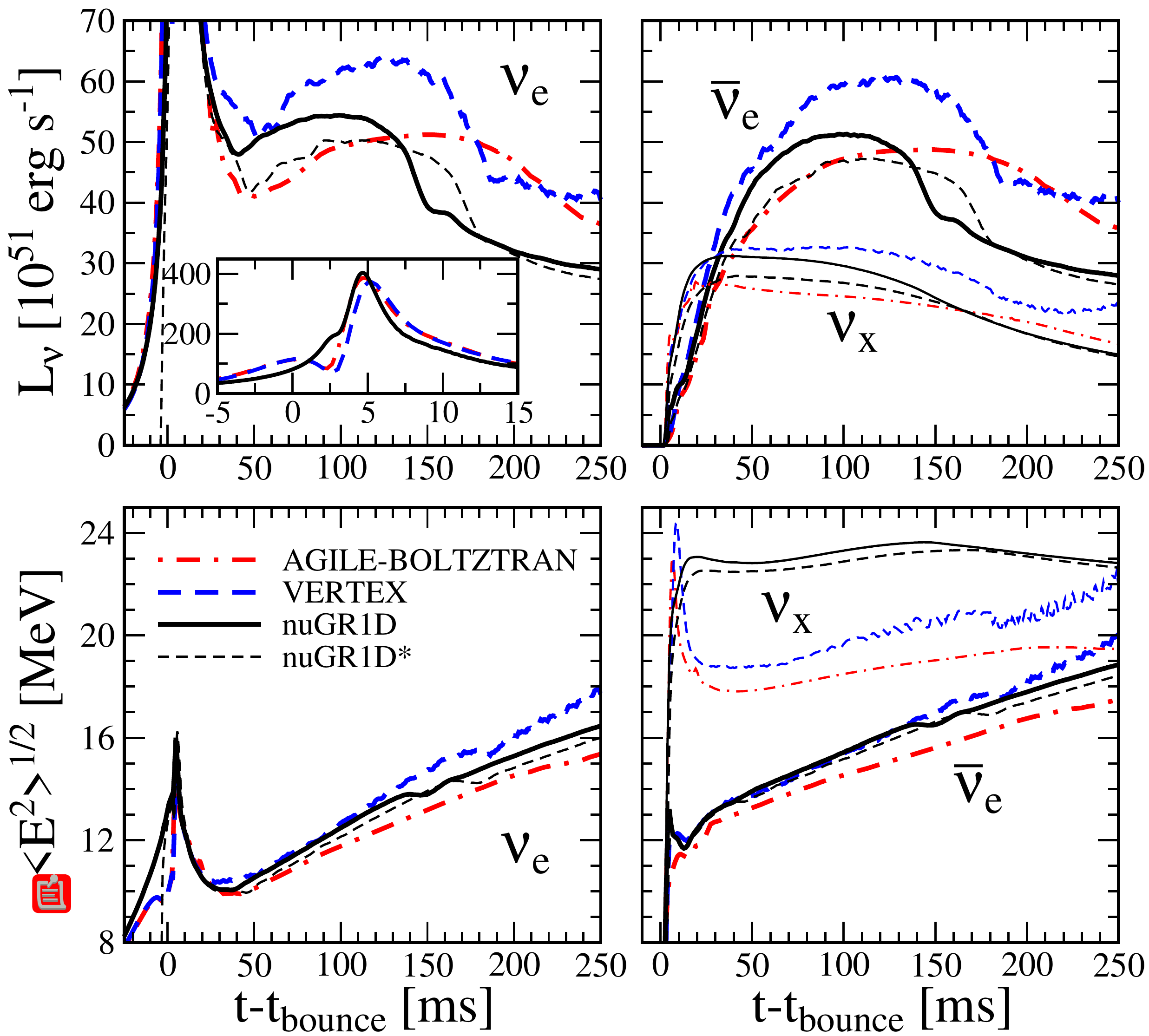}
\caption{Neutrino luminosities and root mean squared energies plotted
  as a function of postbounce time for the $s$15WW95 progenitor.
  These luminosities (top panels) and energies (bottom panels)
  correspond to the comparison study of \cite{liebendoerfer:05}. The
  left panels contain results for $\nu_e$, the right panels show
  $\bar{\nu}_e$ (thick lines) and $\nu_x$ (thin lines) results. The
  inset plot in the upper left panel shows the $\nu_e$ luminosity
  around core bounce. Shown in solid black lines are luminosities and
  root mean squared energies obtained with \code{nuGR1D}.  The blue
  dashed lines and red dashed-dotted lines are the results from
  \cite{liebendoerfer:05} using the \code{VERTEX} code
  \citep{rampp:02} and \mbox{{\it Agile}-\code{BOLTZTRAN}} code
  \citep{liebendoerfer:04}, respectively. The thin dashed black lines
  (labeled as \code{nuGR1D}*) are curves from a test simulation using
  profile data from the \code{VERTEX} simulation 5\,ms before bounce
  as starting data. A detailed discussion of the differences is
  provided in the text.}\label{fig:s15WW95}
\end{figure}

In \fref{fig:s15WW95}, we show the luminosities (top panels) and root
mean squared energies (bottom panels) of three neutrino species:
$\nu_e$ (left panels), and $\bar{\nu}_e$ and $\nu_x$ (right
panels). Both the luminosity and root mean squared energies are
defined in \cite{liebendoerfer:05}, Section~4. The black solid lines
are the results obtained with \code{nuGR1D}, the red dashed-dotted lines are
the predictions of \mbox{{\it Agile}-\code{BOLTZTRAN}}, and the blue
dashed lines are the \code{VERTEX} results. Overall, the agreement is
good, however there are are several systematic differences:

(\emph{i}) The magnitudes of the $\nu_e$ and $\bar{\nu}_e$
luminosities in the early postbounce phase predicted by
\code{nuGR1D} agree well with the \mbox{{\it Agile}-\code{BOLTZTRAN}}
results but they are systematically lower than the \code{VERTEX}
results. This discrepancy, which also exists between \code{VERTEX} and
the \mbox{{\it Agile}-\code{BOLTZTRAN}}, has been further investigated
in \cite{marek:05} and has since been resolved. The updated
\code{VERTEX} code employs an improved general relativistic potential
and gives comparable amplitudes to \mbox{{\it
    Agile}-\code{BOLTZTRAN}}, and hence \code{nuGR1D}.

(\emph{ii}) The time when the silicon--oxygen interface (located at a
baryonic mass coordinate of 1.43\,$M_\odot$ in model $s$15WW95)
accretes through the shock, which is marked by a sharp drop in the
$\nu_e$ and $\bar{\nu}_e$ luminosities, is earlier (at $\sim 140\,$ms)
in our simulations than in the simulations of \cite{liebendoerfer:05}
($\sim 180\,$ms in the \code{VERTEX} simulations\footnote{Such a sharp
drop is not seen in the \mbox{{\it Agile}-\code{BOLTZTRAN}}
results. \cite{liebendoerfer:05} attribute the lack of a sharp drop in
the \mbox{{\it Agile}-\code{BOLTZTRAN}} results to the use of an
adaptive grid, which introduces artificial diffusion and smears out
sharp density features. Nevertheless, the slow decline of the $\nu_e$
and $\bar{\nu}_e$ luminosities begins around the same time as in the
\code{VERTEX} simulations.}). We have carried out a number of tests to
try to understand this difference.  We have varied the zero point of
the internal energy, which affects the relativistic enthalpy entering
the momentum and energy equations and alters the collapse time. We
have attempted different mappings of the presupernova stellar
structure to our Eulerian grid, we have tested variations in
precollapse electron capture by parameterizing $Y_e$ as a function of
density in the collapse phase \citep{liebendoerfer:05fakenu}, and we
have replaced the low density compositions with a pure silicon gas
(which replaces the NSE equivalent of mainly iron and therefore better
represents the low density matter). None of these tests led to a
change of the postbounce time to silicon--oxygen interface accretion
by more than $\sim$\,10\,ms.

Without currently having the modeling technology to test it, we
suspect that the apparent difference may be due to the only other
obvious difference between \code{nuGR1D} and \code{VERTEX}: our lack
of detailed neutrino physics during the collapse phase.  Two facts
lead us to this conclusion. The collapse time discrepancy between the
\code{VERTEX} ($t_\mathrm{collapse} \sim 177$\,ms) and \mbox{{\it
    Agile}-\code{BOLTZTRAN}} ($t_\mathrm{collapse} \sim 172$\,ms) and
\code{nuGR1D} ($t_\mathrm{collapse} \sim 224$\,ms) can be explained by
the lack of inelastic scattering. \cite{liebendoerfer:11pc} gives a
$t_\mathrm{collapse} \sim 213$\,ms in \mbox{{\it
    Agile}-\code{BOLTZTRAN}} when neutrino-electron inelastic
scattering is neglected. This collapse time difference, $\sim$40\,ms,
is the difference in the silicon-oxygen interface accretion
times. Also, when we map in the \code{VERTEX} profiles from
\cite{liebendoerfer:05} at 5\,ms before bounce to \code{nuGR1D} and
continue the evolution we reproduce the postbounce time that the
silicon-oxygen interface reaches the shock. We show the result of this
simulation with a thin dashed curve, marked as nuGR1D*, in
\fref{fig:s15WW95}. It is worth noting that there may be additional
differences due to the simplistic treatment of the EOS at low
densities.  Below a density of $6\times 10^{7}$\,g\,cm$^{-3}$,
\code{VERTEX} replaces the nuclear statistical equilibrium (NSE) EOS
with an EOS that specifically depends on the composition of the
matter. \mbox{{\it Agile}-\code{BOLTZTRAN}} assumes all matter below
this density is silicon. Both have some treatment of nuclear
burning. In \code{GR1D}, as discussed in \cite{oconnor:10}, we assume
NSE compositions from the nuclear EOS. At densities below the validity
regime of the nuclear EOS, we take the compositions at the lowest
density point from the nuclear EOS and use the Timmes EOS
\citep{timmes:99}. For reference, the initial density of the
silicon--oxygen interface is 0.4--1.0$\times 10^7$\,g\,cm$^{-3}$,
where the range represents the extent in density space.

(\emph{iii}) The $\nu_e$ and $\bar{\nu}_e$ root mean squared energies,
predicted by \code{nuGR1D} agree very well with the Boltzmann
transport results during the postbounce phase. The difference seen in
the $\nu_x$ root mean squared energy is similar to that observed by
\cite{thompson:02phd} and \cite{lentz:12b} when investigating the
effects of inelastic neutrino--electron scattering. In the postbounce
evolution, this interaction is expected to predominately affect the
$\nu_x$ neutrino. We currently ignore this process in \code{nuGR1D}
and note that the $\nu_x$ luminosity predicted by \code{nuGR1D} still
agrees well with the full Boltzmann results.

(\emph{iv}) Another difference between the evolution in \code{nuGR1D}
and the full Boltzmann transport results arises in the collapse phase.
The lack of velocity terms and inelastic $\nu_e - e^-$ scattering
significantly effects the composition of the inner core.  In
simulations with inelastic $\nu_e - e^-$ scattering, neutrinos from
electron capture on free protons down-scatter off of electrons to
lower energies.  Since the optical depth is lower, these neutrinos can
then escape, deleptonizing the core. In our simulations, these
high-energy neutrinos cannot down-scatter and therefore cannot
escape. Deleptonization is suppressed until later phases, when central
density and temperature are higher. The lack of velocity-dependent
terms delays full trapping by neutrino advection that would normally
begin to occur at $\rho \gtrsim 1\times10^{12}$\,g\,cm$^{-3}$ until
nuclear densities, allowing for further deleptonization. At bounce the
central value of $Y_e$ are $\sim$\,0.22 compared to $\sim$0.29 in
\code{VERTEX} and \mbox{{\it Agile}-\code{BOLTZTRAN}}.  The root mean
squared $\nu_e$ energies predicted by \code{nuGR1D} are higher during
the prebounce phase, because the neutrinos do not experience the
down-scattering via inelastic neutrino--electron scattering.

Overall, we find that differences in the inner core structure and
composition at bounce do not strongly present themselves in the
neutrino signal after the collapse phase. Therefore, for this study,
we find the current version of \code{nuGR1D} to be acceptable, since
our primary focus is the neutrino signal of the preexplosion accretion
phase.  We do note, however, that the lack of energy-coupling terms in
our transport can cause qualitative differences near black hole
formation.  When energy-coupling terms, such as gravitational
redshift, are included, then the $\nu_e$ and $\bar{\nu}_e$
luminosities drop off at times very close to black hole formation, as
seen in \cite{fischer:09a}.  This effect can be captured with
\code{GR1D}'s leakage scheme in which it is trivial to include
redshift terms \citep{oconnor:10}. Capturing redshift in an
energy-dependent transport scheme requires energy-group coupling. We
have also compared our M1 scheme to the results of \cite{fischer:09a}
for the 40\,$M_\odot$ model from \cite{ww:95} using the LS180 EOS.  In
this model a black hole forms within 500\,ms of bounce. We find
differences in the $\nu_e$ and $\bar{\nu}_e$ luminosities of
$\sim$\,10--20\% in the last $\sim$50\,ms, but good agreement in the
early postbounce phase.

\section{Results}
\label{sec:results}

\begin{deluxetable}{rcccccccccccc}
  \tablecolumns{5} \tablewidth{0pc} \tablecaption{Key Neutrino
    Quantities} \tablehead{ Model & $\xi_{1.75}$ & $E_\mathrm{400ms}^{\nu_e}$/$E_\mathrm{400ms}^{\bar{\nu}_e}$/$E_\mathrm{400ms}^{\nu_x}$& $N_\mathrm{200ms}^\mathrm{ibd}$& $N_\mathrm{400ms}^\mathrm{ibd}$\vspace*{0.1cm}\\
&&&{\scriptsize LS220/HShen}&{\scriptsize LS220/HShen}\\
&&[B]&[$10^3$]&[$10^3$] }
\startdata
$s$12 & 0.235 & 19.24 / 14.19 / \phm{0}7.73 & 1.02 / 0.92 & 2.13 / 1.78 \\
$s$13 & 0.383 & 22.18 / 16.58 / \phm{0}8.76 & 1.25 / 1.09 & 2.53 / 2.07 \\
$s$14 & 0.537 & 25.19 / 19.35 / \phm{0}9.24 & 1.36 / 1.20 & 3.06 / 2.49 \\
$s$15 & 0.580 & 25.51 / 19.59 / \phm{0}9.17 & 1.30 / 1.16 & 3.13 / 2.59 \\
$s$16 & 0.338 & 18.91 / 13.72 / \phm{0}8.20 & 1.11 / 0.95 & 2.00 / 1.68 \\
$s$17 & 0.383 & 19.93 / 14.54 / \phm{0}8.57 & 1.20 / 1.02 & 2.13 / 1.78 \\
$s$18 & 0.738 & 28.66 / 22.26 / 10.30 & 1.55 / 1.36 & 3.62 / 2.92 \\
$s$19 & 0.544 & 23.36 / 17.49 / \phm{0}9.37 & 1.43 / 1.23 & 2.67 / 2.18 \\
$s$20 & 0.944 & 29.39 / 22.70 / 11.08 & 1.79 / 1.55 & 3.64 / 2.91 \\
$s$21 & 0.325 & 18.48 / 13.41 / \phm{0}8.09 & 1.04 / 0.91 & 1.95 / 1.65 \\
$s$22 & 0.972 & 30.06 / 23.29 / 11.26 & 1.82 / 1.58 & 3.76 / 3.00 \\
$s$23 & 1.256 & 42.00 / 33.88 / 14.68 & 2.29 / 2.00 & 5.99 / 4.67 \\
$s$24 & 1.167 & 39.14 / 31.35 / 13.66 & 2.16 / 1.89 & 5.44 / 4.24 \\
$s$25 & 1.040 & 32.30 / 25.28 / 11.87 & 1.93 / 1.68 & 4.15 / 3.28 \\
$s$26 & 0.727 & 24.97 / 18.84 / 10.10 & 1.59 / 1.38 & 2.88 / 2.33 \\
$s$27 & 0.783 & 25.67 / 19.45 / 10.21 & 1.61 / 1.40 & 3.01 / 2.42 \\
$s$28 & 0.640 & 25.86 / 19.83 / \phm{0}9.54 & 1.36 / 1.21 & 3.16 / 2.62 \\
$s$29 & 0.556 & 23.24 / 17.38 / \phm{0}9.43 & 1.45 / 1.22 & 2.64 / 2.17 \\
$s$30 & 0.760 & 26.61 / 20.30 / 10.19 & 1.60 / 1.40 & 3.20 / 2.58 \\
$s$31 & 0.687 & 25.23 / 19.09 / \phm{0}9.95 & 1.56 / 1.35 & 2.96 / 2.40 \\
$s$32 & 0.883 & 28.10 / 21.58 / 10.70 & 1.71 / 1.49 & 3.43 / 2.75 \\
$s$33 & 0.965 & 30.36 / 23.57 / 11.28 & 1.82 / 1.58 & 3.82 / 3.04 \\
$s$35 & 1.129 & 35.93 / 28.50 / 12.77 & 2.06 / 1.79 & 4.83 / 3.79 \\
$s$40 & 1.328 & 48.32 / 39.45 / 17.51 & 2.49 / 2.21 & 7.23 / 5.75 \\
$s$45 & 1.300 & 46.42 / 37.81 / 16.56 & 2.42 / 2.14 & 6.86 / 5.44 \\
$s$50 & 0.701 & 25.60 / 19.42 / 10.00 & 1.57 / 1.32 & 3.02 / 2.45 \\
$s$55 & 0.577 & 23.56 / 17.65 / \phm{0}9.52 & 1.48 / 1.25 & 2.69 / 2.20 \\
$s$60 & 0.461 & 21.52 / 15.88 / \phm{0}9.07 & 1.33 / 1.12 & 2.36 / 1.95 \\
$s$70 & 0.755 & 25.83 / 19.58 / 10.17 & 1.60 / 1.39 & 3.04 / 2.46 \\
$s$80 & 0.591 & 23.75 / 17.80 / \phm{0}9.63 & 1.48 / 1.26 & 2.71 / 2.22 \\
$s$100 & 0.792 & 29.90 / 23.33 / 10.59 & 1.61 / 1.42 & 3.82 / 3.07 \\
$s$120 & 0.474 & 22.31 / 16.64 / \phm{0}9.02 & 1.33 / 1.15 & 2.52 / 2.07 \\
\enddata

\tablecomments{For each model in the \cite{woosley:07} model set we
  show $\xi_{1.75}$, the cumulative emitted neutrino energy in
  $\nu_e$, $\bar{\nu}_e$ and a single $\nu_x$ at 400\,ms after bounce.
  The numbers correspond to the models run with the LS220 EOS.  We
  also present, for each model, the estimated number of IBD
  interactions in a Super-Kamiokande-like detector at 200 and 400\,ms
  after bounce for a supernova at a fiducial galactic distance of
  10\,kpc for both EOS.}  \label{tab:results}
\end{deluxetable}

\subsection{Trends in the Neutrino Observables}
\label{sec:trends}

\begin{figure*}[ht]
\centering
\includegraphics[width=0.97\textwidth]{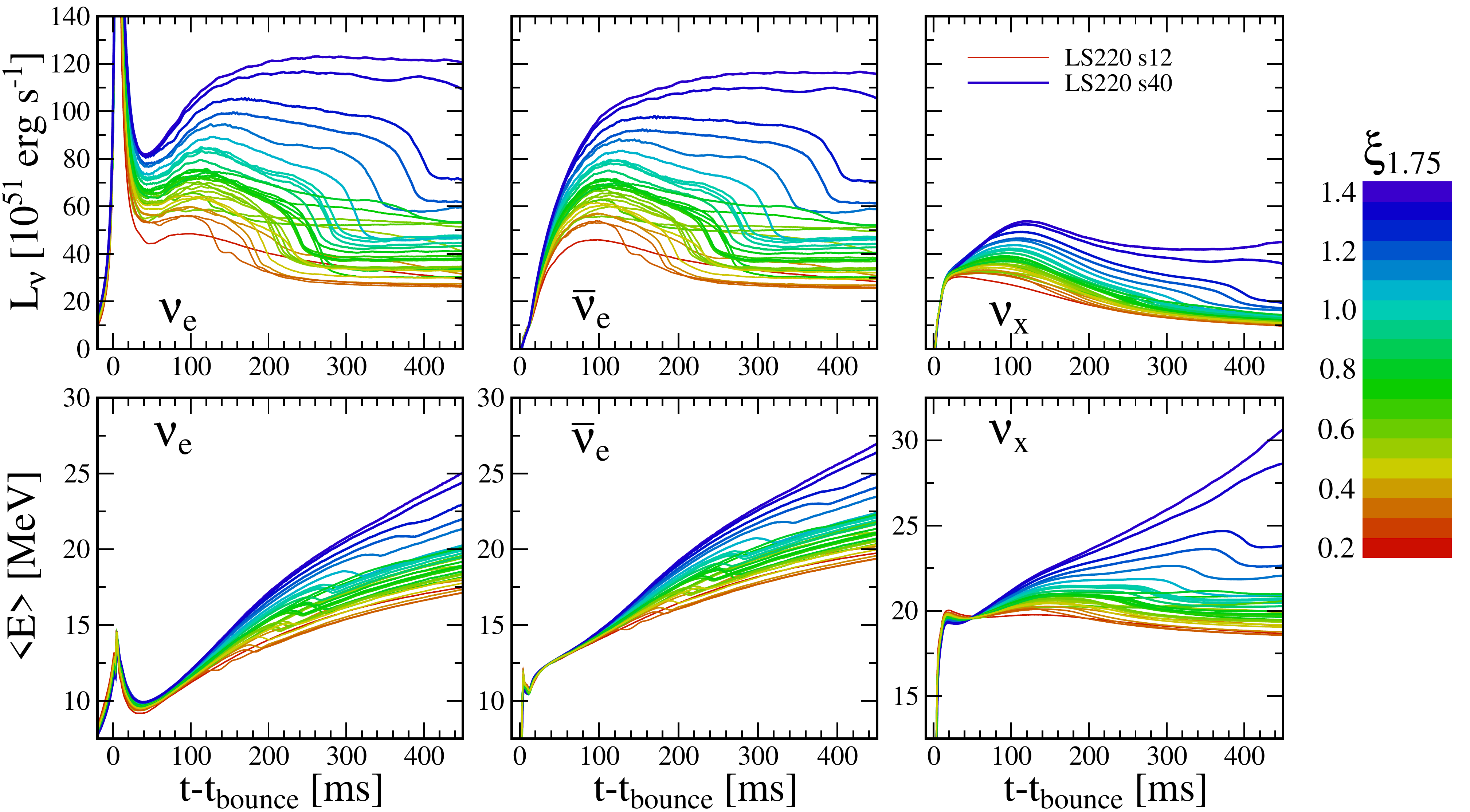}\\
\includegraphics[width=0.97\textwidth]{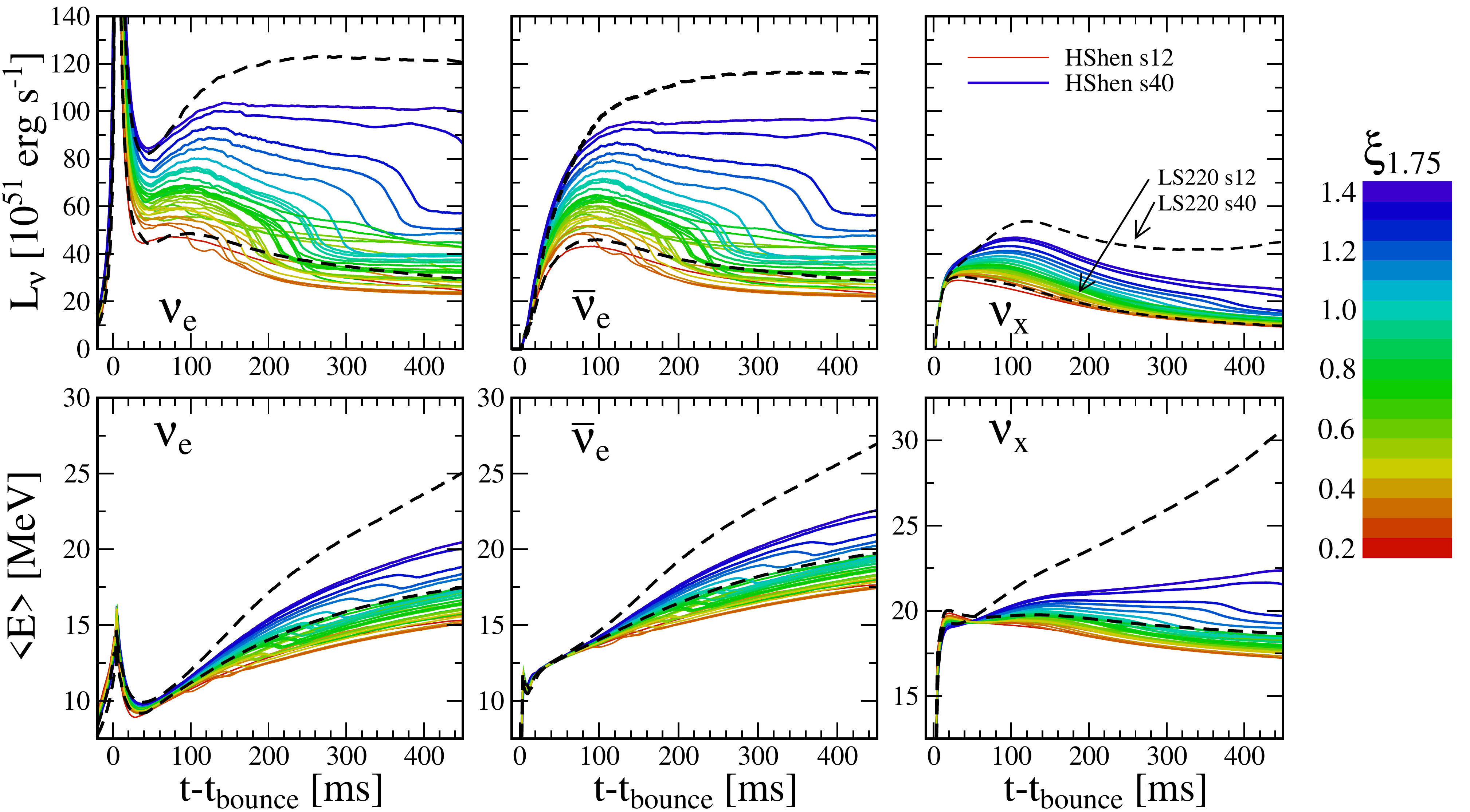}
\caption{Neutrino luminosities (top panels) and average energies
  (bottom panels) plotted as a function of postbounce time for all 32
  models of \cite{woosley:07}. The top set of panels shows results
  obtained with the LS220 EOS. The bottom panel shows the same for the
  HShen EOS, but includes, for reference, two LS220 models: $s$12WH07
  and $s$40WH07. The left, center, and right panels show results for
  $\nu_e$, $\bar{\nu}_e$, and $\nu_x$, respectively. The curves are
  color- and line-weight-coded with increasing compactness
  ($\xi_{1.75}$), the mapping from color to compactness parameter is
  shown on the right. There is a clear trend in all luminosities and
  average energies with compactness parameter.  The progenitor with
  the highest compactness, $s$40WH07, forms a black hole at 503\,ms
  after bounce. None of these models explode, but the onset of an
  explosion in any of these models may lead to a sudden deep drop
  (strongest for $\nu_e$ and $\bar{\nu}_e$) in the luminosities and
  average energies \citep{fischer:10}, although this is likely
  suppressed by multidimensional effects. The smaller drop observed
  for most models models here is due to the sudden decrease of the
  accretion rate when the silicon--oxygen interface reaches the
  stalled shock.}\label{fig:lums_and_aveen}
\end{figure*}

We perform core collapse and early postbounce evolutions of all 32
models introduced in \sref{sec:models} using both the LS220 and the
HShen EOS.  In \fref{fig:lums_and_aveen} we present the three neutrino
luminosities and average energies for each model and EOS as a function
of postbounce time. We do not expect a clear trend in the neutrino
observables with ZAMS mass. However, we do expect trends based on the
presupernova structure of the star, which is well encapsulated by the
compactness parameter introduced in \S\ref{sec:models}. In the top set
of panels we show simulations run using the LS220 EOS. In the bottom
panels, simulations performed with the HShen EOS are shown. To
highlight that there is indeed a trend with presupernova structure, we
color-code individual models according to their compactness
parameter. The mapping between line color and $\xi_{1.75}$ is provided
on the right.  To more directly highlight the EOS dependence, we
include the luminosity and average energies of two models run with the
LS220 EOS in the HShen EOS panels with thick dashed lines.  These
models, $s$12WH07 and $s$40WH07, have the lowest and highest
compactness parameter in our model set, respectively.

We find that there is little variation in the peak luminosity of the
$\nu_e$ neutronization burst signal. For all 32 models simulated using
the LS220 (HShen) EOS, the peak amplitude varies by less than 3\%
(5\%) from the average. This reflects the universal nature of the
collapse of the inner core \citep{liebendoerfer:02b}. After the
neutronization burst, the postbounce luminosities of all species
increase systematically with increasing compactness parameter.  Models
with higher $\xi_{1.75}$ have higher temperatures throughout the
protoneutron star \citep{oconnor:11}. This increases the diffusive
neutrino luminosity and is best seen in the $\nu_x$ luminosities.  The
postbounce accretion rate also increases with the compactness
parameter. Higher accretion rates, and the deeper gravitational
potential due to the higher protoneutron star mass, increase the
accretion luminosity, which is most directly reflected in the $\nu_e$
and $\bar{\nu}_e$ signals.  After roughly 100\,ms, the average energy
of the emitted neutrinos also shows an increasing trend with the
compactness parameter.  The matter temperature at the neutrinosphere
is higher in models with larger compactness, therefore a higher
average neutrino energy is observed at infinity.

The neutrino luminosities and average energies from simulations using
the HShen EOS are systematically lower than the luminosities and
average energies from simulations of the same model run with the LS220
EOS.  This is clearly seen in the bottom set of panels in
\fref{fig:lums_and_aveen}: models $s$12WH07-LS220 and $s$40WH07-LS220
have luminosities and average energies that are comparable to or
larger than in the corresponding HShen models.  For a fixed accretion
rate (or fixed progenitor model), the location of the neutrinosphere
of each species influences the emitted luminosities and spectra. In
models evolved with the stiff HShen EOS, the neutrinospheres are
located systematically at larger radii and lower matter temperatures
than in models run with the softer LS220 EOS. For example, in the
$s$12WH07 simulations, the Rosseland-mean $\nu_e$ neutrinospheres at
200\,ms after bounce have radii and temperatures of $\sim$\,35.3\,km
and $\sim$\,4.86\,MeV; and $\sim$\,39.5\,km and $\sim$\,4.46\,MeV for
the LS220 and the HShen EOS, respectively.  The larger neutrinosphere
radii are responsible for the lower accretion luminosity since the
latter is set essentially by the product of the mass accretion rate
and the gravitational potential at the protoneutron star surface
\citep{liebendoerfer:02b}. The latter is located at larger radii in
simulations using the HShen EOS.  The differences in the
neutrinosphere radii and temperatures between the LS220 and HShen EOS
also give an explanation for the systematically lower average neutrino
energies seen in the HShen simulations.  Matter at larger radii has
been compressed less and therefore is cooler. This leads to average
neutrino energies that can be up to 5\,MeV lower for the HShen EOS
than for the LS220 EOS for the same progenitor model (see
\fref{fig:lums_and_aveen}).  The difference in the neutrino
luminosities and average energies between the two EOS is largest for
models with $\xi_{1.75} \gtrsim 1.2$, in which the high accretion
rates lead to the accumulation of $\sim$\,2\,$M_\odot$ of material
inside the shock within $\sim$\,200--300\,ms of bounce.  In the case
of the LS220 EOS, this leads to very high temperatures throughout the
protoneutron star as it becomes more and more compact and closer to
gravitational collapse to a black hole. In our simulations, the most
compact model $s$40WH07 ($\xi_{1.75} = 1.33$, $\xi_{2.5} = 0.59$)
forms a black hole 503\,ms after bounce.  The slightly less compact
model $s$45WH07 ($\xi_{1.75} = 1.30$, $\xi_{2.5} = 0.55$) forms a
black hole 563\,ms after bounce. The high temperatures present in the
LS220 simulations at these times will not be obtained until postbounce
times $\gtrsim 1\,\mathrm{s}$ in models using the HShen
EOS\footnote{While we do not follow these models to black hole
  formation in our current study, in \cite{oconnor:11} we found that
  the black hole formation times of the $s$40WH07 and $s$45WH07 models
  are $\sim$1.3\,s and $\sim$1.4\,s, respectively, when using the
  HShen EOS.}.

\begin{figure*}[ht]
\centering 
\includegraphics[trim=0cm 10.3cm 0cm 0cm, clip=true,width=0.97\textwidth]{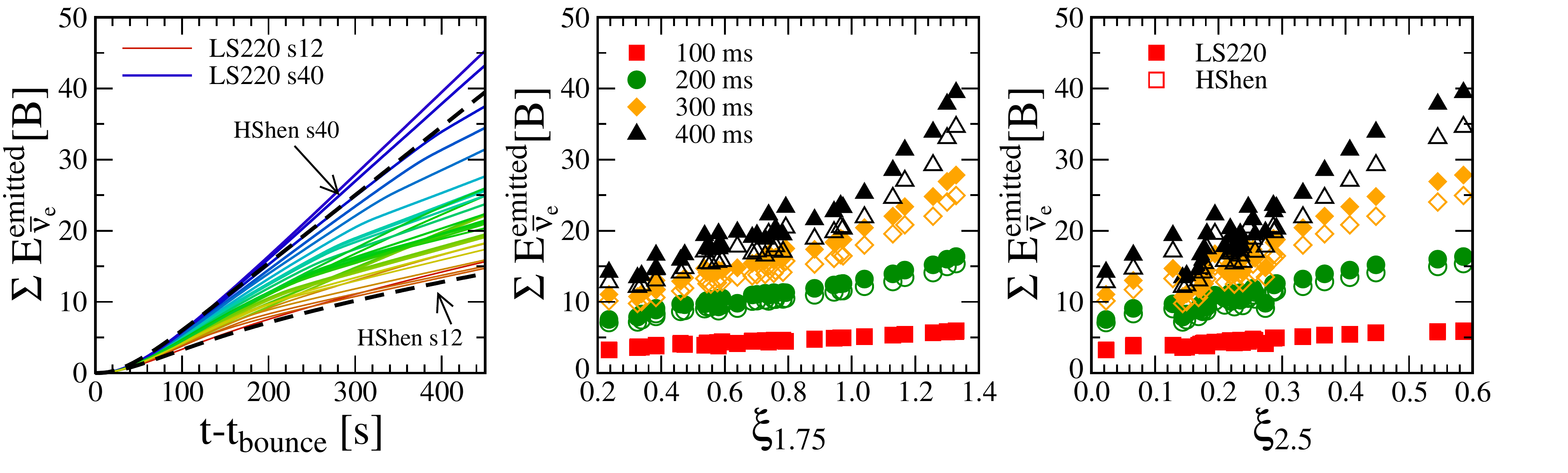}
\caption{Cumulative emitted electron antineutrino energy. We show the
  cumulative totals via several methods.  In the left panel we show
  time series data for each of the 32 models of \cite{woosley:07} 
  run with the LS220 EOS.  The color coding corresponds to the value of
  $\xi_{1.75}$ and is given in \fref{fig:lums_and_aveen}.  This
  cumulative energy is the time integral of the top-center panel of
  \fref{fig:lums_and_aveen}.  The dashed black lines correspond to
  models $s$12WH07 and $s$40WH07 run with the HShen EOS. In the
  center and right panels we present the emitted energy at select
  postbounce times, for each model and EOS, plotted versus
  the compactness $\xi_{1.75}$ (center panel) and $\xi_{2.5}$ (right
  panel) of the model. \label{fig:detectability}}
\end{figure*}

Most neutrino detectors are most sensitive to the electron
antineutrino luminosity through the dominant IBD interaction,
$\bar{\nu}_e + p \to n + e^+$.  In the left panel of
\fref{fig:detectability}, we consider the cumulative emitted
$\bar{\nu}_e$ energy for each model using the LS220 EOS. We color code
the models based on their compactness parameter and include two
 reference models that use the HShen EOS, model $s$12WH07 and model
$s$40WH07.  The graphs shown in this panel are the integral of the
graphs shown in the top center panel of \fref{fig:lums_and_aveen}.  It
is obvious that the cumulative amount of $\bar{\nu}_e$ energy emitted
during the preexplosion phase strongly correlates with the compactness
of the progenitor model. For example, the amount of emitted
$\bar{\nu}_e$ energy from model $s$40WH07 ($\xi_{1.75} = 1.33$) is
\emph{always} between two and three times of that of model $s$12WH07
($\xi_{1.75} = 0.24$). We make this point more quantitative in the
center and right panels of \fref{fig:detectability}.  In the center
(right) panel we plot the cumulative emitted $\bar{\nu}_e$ energy at
100, 200, 300, and 400\,ms after bounce for both EOS as a function of
$\xi_{1.75}$ ($\xi_{2.5}$).  For reference, we present a subset of
these numbers in \tref{tab:results}. We see a very clear correlation
that depends only weakly on the chosen EOS. Note, however, that for
models with small compactness parameter ($\xi_{1.75} \lesssim 0.8$),
the correlation between the total emitted $\bar{\nu}_e$ energy and the
compactness parameter is not as strong after 400\,ms of postbounce
evolution.  Comparing the center and right panels justifies our choice
of $\xi_{1.75}$ over $\xi_{2.5}$ as explained in \sref{sec:models}.

The onset of an explosion will break the correlation observed in
\fref{fig:detectability}.  Once it is launched, the accretion
luminosity effectively turns off and only the diffusion luminosity
remains.  One also expects this diffusion luminosity to show a
correlation with the compactness of the progenitor, since the remnant
protoneutron star's thermodynamic conditions, such as the central
entropy and its mass, are essentially set by the presupernova
structure. However, it is currently unclear whether one should
obtain a correlation between explosion time and the compactness
parameter. Clarification will require a more complete understanding of
the core-collapse supernova explosion mechanism and may require
extensive parameter studies with fully self-consistent
three-dimensional simulations.

\subsection{Detectability}
\label{sec:detect}

\begin{figure}[]
\centering
\includegraphics[width=0.97\columnwidth]{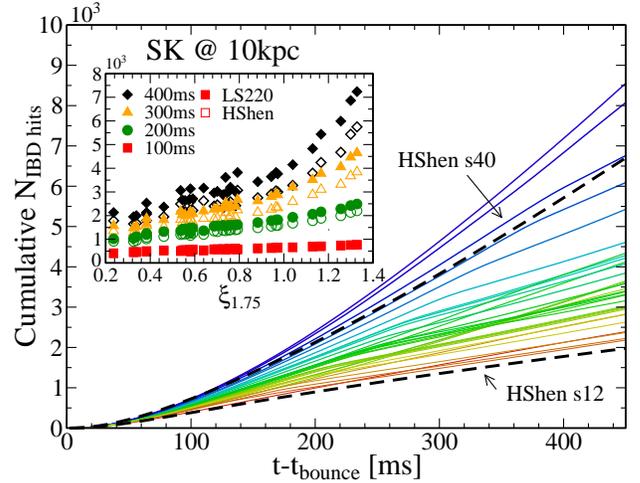}
\caption{Cumulative IBD interactions in a Super-Kamiokande-like water
  Cherenkov detector at a fiducial galactic distance of 10\,kpc versus
  postbounce time.  We use the \code{SNOwGLoBES} package to determine
  the integrated IBD interaction rate in a 32\,kT water Cherenkov detector
  at 10\,kpc.  The color coding corresponds to the value of $\xi_{1.75}$
  and is provided in \fref{fig:lums_and_aveen}. The dashed lines are results
  for models $s$12WH07 and  $s$40WH07 run with the HShen EOS. In the
  inset we show the cumulative IBD interactions as a function of
  $\xi_{1.75}$ for each model and EOS at four postbounce times: 100,
  200, 300, and 400\,ms.}\label{fig:cumul_nhits}
\end{figure}

We use the publicly available software \code{SNOwGLoBES}
\footnote{available at \url{http://www.phy.duke.edu/\textasciitilde
    schol/snowglobes}} to predict the neutrino signal observed in
Earth-based neutrino detectors.  \code{SNOwGLoBES}
\citep{snowglobes:may12,scholberg:12}, which in turn relies on
\code{GLoBES} \citep{huber:04,huber:07}, is a set of routines that
compute the interaction rates of supernova neutrinos in user-specified
detector configurations. Variables include detector material (e.g.,
water, scintillator, argon, and lead), detector volume, detector
response functions, and a host of relevant neutrino interactions. For
this investigation, we consider only IBD interactions in a water
Cherenkov detector. We choose a detector mass of 32\,kT, the mass of
water in Super-Kamiokande that is sensitive to core-collapse supernova
neutrinos \citep{scholberg:12}.  For reference, we use the
\code{wc100kt30prct} smearing rates and efficiencies provided with
\code{SNOwGLoBES}.  We construct \code{SNOwGLoBES} initial fluence
data from our simulations binned in 5\,ms intervals. We provide these
energy-dependent fluences at {\tt
  \url{http://www.stellarcollapse.org/M1prog}} for all models,
neutrino species, and both EOS in 5\,ms intervals up to 450\,ms after
bounce. We assume a fiducial galactic supernova distance of 10\,kpc.
In \fref{fig:cumul_nhits}, we show the cumulative number of
interactions for each model run with the LS220 EOS and, for reference,
two models run with the HShen EOS.  The lines are color-coded
according to compactness parameter. Note that the vertical scale in
this figure is in thousands of interactions.  We note that our answers
agree with the total number of expected detected interactions in
Super-Kamiokande from a galactic core-collapse supernova at 10\,kpc,
which is estimated to be $\sim$\,7000 \citep{scholberg:12}.  To arrive
at this number from our results, consider the lowest ZAMS mass
progenitor in our model set, model $s$12WH07.  After 450\,ms of
evolution, 15\,B of $\bar{\nu}_e$ energy has been radiated
(\fref{fig:detectability}), which corresponds to 2000 IBD detected
interactions (\fref{fig:cumul_nhits}).  For 50\,B of released energy
($\sim$\,1/6 of 300\,B, the fiducial energy released in neutrinos over
the entire cooling phase), one would then expect $\sim$\,7000 detected
interactions. However, as is clear from \fref{fig:cumul_nhits}, the
number of detected interactions from the next galactic supernova may
be higher than this fiducial number. More importantly, the rate of
interactions in the preexplosion phase will give us detailed
information on the progenitor core structure.

In order to more directly quantify the differences between variations
in progenitor compactness and variations in the nuclear EOS, we plot
in the inset of \fref{fig:cumul_nhits} the number of expected IBD
detected interactions in a Super-Kamiokande-like water Cherenkov
detector at various postbounce times versus $\xi_{1.75}$.  There is a
well defined trend: The number of IBD interactions detected in the
first 100, 200, 300, and 400\,ms increases with the compactness
parameter of the models.  For reference, we include the expected
number of interactions at 200 and 400\,ms for both EOS in
\tref{tab:results}. We find that the EOS dependence of the expected
number of interactions is similar to the EOS dependence of the total
emitted $\bar{\nu}_e$ energy: the HShen EOS leads to a lower number of
interactions (compare the inset of \fref{fig:cumul_nhits} to the
center panel of \fref{fig:detectability}). The dependence on EOS is
somewhat stronger here, since the the lower average $\bar{\nu}_e$
energy predicted from stiffer EOS translates into a reduced cross
section in Earth-based detectors.  In addition to the total number of
interactions, a water Cherenkov detector measures individual energies, and
thus, allows for the reconstruction of the cumulative emitted
$\bar{\nu}_e$ energy over time.  This reconstruction will depend on
the detector's response function and efficiency.

An additional independent path to experimentally probing the inner
structure of the progenitor is via the total neutrino energy emitted
in all species over the first 10s of seconds after the initial
collapse.  This method requires a measurement of the total fluence of
neutrinos of all species, not just electron antineutrinos.  Examples
of neutrino interactions capable of relaying such information are the
mono-energetic de-excitation of a neutral-current neutrino-excitation
of $^{12}$C \citep{scholberg:12} or neutrino-proton elastic scattering
interactions \citep{dasgupta:11}.  Such measurements would require
good energy resolution, a significant source of carbon and/or a low
energy threshold, for example, a liquid scintillator neutrino
detector. We note that even with a liquid scintillator detector, the
dominant neutrino interaction is still IBD \citep{scholberg:12}.

If such a measurement was made, and there is not a significant amount
of rotation (see the discussion on rotation in \sref{sec:degen}), one
can immediately infer the gravitational binding energy of the remnant,
since neutrinos carry away the vast majority ($\sim 99\%$\footnote{The
  remaining $\sim$1\% of the energy is predominantly shared among the
  kinetic energy of the explosion, the original binding energy of the
  unbound stellar mantle, and the binding energy of the iron core at
  the onset of collapse.}) of the gravitational binding energy. For
typical nuclear EOS like the ones considered here, this results in a
one-to-one mapping of the released gravitational binding energy to the
baryonic mass of the remnant, and, hence, the gravitational mass of the
remnant. This is most easily seen by fitting the gravitational binding
energy of a cold (T=0.1\,MeV), neutrino-less $\beta$-equilibrium,
non-rotating neutron star to its baryonic mass.  From cold neutron
star TOV solutions using the LS220 EOS one can obtain an empirical fit
to better than 3\% above a baryonic mass of 1.15\,$M_\odot$,
\begin{equation}
E_\mathrm{binding} \sim \,1.12\times 10^{53} (M_\mathrm{bary} /
M_\odot)^2\, \mathrm{ergs}\,.\label{eq:ebind_ls220}
\end{equation}  
A similar fit for the HShen EOS gives,
\begin{equation}
E_\mathrm{binding} \sim 9.78\times10^{52} (M_\mathrm{bary} /
M_\odot)^2\, \mathrm{ergs}\,,\label{eq:ebind_hshen}
\end{equation}
and is accurate to 5\% above baryonic masses of 1.15\,$M_\odot$.
Below $M_\mathrm{bary}=1.15\,M_\odot$, the empirical quadratic fit is
not as accurate. However, all models considered here reach a baryonic
protoneutron star mass of 1.15\,$M_\odot$ within $\sim$\,10\,ms of
bounce. Hence, we believe that the above fits are acceptable for the
iron-core core collapse events considered here.

\begin{figure}[t]
\centering
\includegraphics[width=0.97\columnwidth]{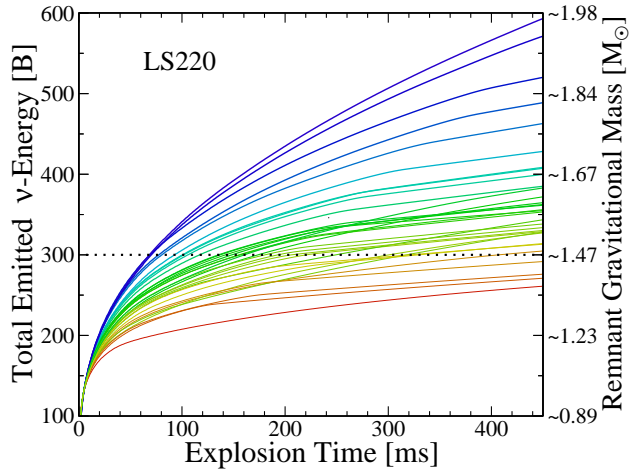}
\caption{Total emitted neutrino energy estimated from the enclosed
  baryonic mass as a function of explosion time for the LS220 EOS.  A
  measurement of the total emitted neutrino energy and an estimate of
  the explosion time constrains progenitor structure.  For reference,
  we provide the cold neutron star gravitational mass associated with
  the released binding energy on the right ordinate. This figure is
  constructed using a fit of the gravitational binding energy of a
  cold neutron star to its baryonic mass, $E^\mathrm{total}_{\nu}
  \sim\,1.12\times10^{53}(M_\mathrm{bary}/M_\odot)^2\,\mathrm{ergs}$,
  and the baryonic mass enclosed in the shock at any given time.  This
  defines the explosion time to be, in a Lagrangian sense, the time at
  which the outermost final neutron star mass element accretes through
  the shock. The color coding corresponds to $\xi_{1.75}$, the color
  coding is provided in
  \fref{fig:lums_and_aveen}. }\label{fig:total_emitted_energy}
\end{figure}

\begin{figure}[ht]
\centering
\includegraphics[width=0.97\columnwidth]{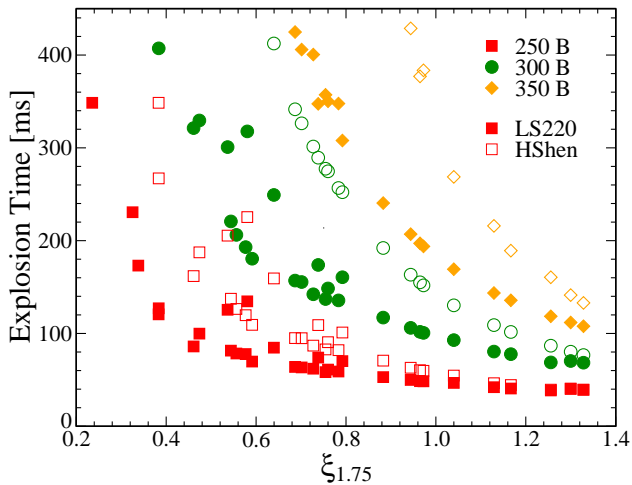}
\caption{Explosion time for various progenitors and EOS, under the
  constraint of a fixed amount of released neutrino energy.  We show
  three values of the total released energy, 250, 300, and
  350\,B. Progenitors with high compactness must explode at earlier
  times to achieve the same total amount of released energy in
  neutrinos when compared to a progenitor with a lower
  compactness. The strong EOS dependence is due to the different
  binding energies resulting from the two EOS. The LS220 EOS leads to
  a more compact neutron star, releasing $\sim\,$14\% energy for a
  given baryonic mass.}\label{fig:explosion_time_vs_xi175}
\end{figure}

We now make the assumption that an explosion launched at a particular
postbounce time will result in a neutron star remnant with a baryonic
mass equal to the baryonic mass that has accreted through the shock up
until the time of the explosion. This neglects any late-time fallback
of material onto the protoneutron star which would lead to additional
neutrino emission. \cite{fryer:09} and \cite{ugliano:12} predict
fallback masses $\lesssim$5-10\% of the initial protoneutron star
remnant mass. We also neglect any asymmetric mass accretion which may
occur in the early explosion phase. In
\fref{fig:total_emitted_energy}, we convert the baryonic mass enclosed
by the shock to the \emph{total} emitted neutrino energy using
\esref{eq:ebind_ls220} and \ref{eq:ebind_hshen}.  We plot this for all
progenitor models (run with the LS220 EOS) as a function of the
hypothetical time of explosion.  As a concrete example, consider the
situation where 300\,B (shown as the dashed line in
\fref{fig:total_emitted_energy}) of total neutrino energy was
observational inferred. This could correspond ($i$) to a progenitor
with a high compactness parameter that exploded at an early time,
e.g., model $s$40WH07 at 70\,ms or ($ii$) to a low progenitor with low
compactness parameter that exploded at late times, e.g., model
$s$12WH07 at 400\,ms. If we have an estimate of the explosion time,
e.g., via characteristic features in the neutrino observables, then we can
use the combined measurement to probe the progenitor core
structure. This is further quantified in
\fref{fig:explosion_time_vs_xi175}, where we choose three total
emitted neutrino energies, 250, 300, and 350\,B, and determine the
time at which the explosion must have been launched for a given
compactness and the respective total emitted energy.  We plot this
explosion time versus $\xi_{1.75}$ for all models and both EOS. In
general, for a fixed total emitted energy, as the compactness
parameter of the progenitor increases, the explosion time must
decrease. For progenitors with high $\xi_{1.75}$ ($\gtrsim 0.8$),
there is a clear mapping between the explosion time and the
compactness parameter, given a specific total emitted neutrino energy
and an EOS. As was the case for the total IBD rates, there is some
scatter at low $\xi_{1.75}$ ($\lesssim 0.8$), but there is still an
overall trend.

\fref{fig:explosion_time_vs_xi175} shows that there is a very strong
EOS dependence. While the baryonic mass inside the shock as a function
of postbounce time does not vary strongly with EOS, the gravitational
binding energy released does.  Based on the above empirical fits to
the gravitational binding energy, the LS220 EOS leads to $\sim\,14\%$
more energy release than the HShen EOS for the same baryonic
mass. This method of determining the compactness parameter (by
combining an estimate of the explosion time and the total emitted
neutrino energy) can be used together with the methods described above
(using the postbounce, preexplosion IBD rates) as a consistency check,
or to break degeneracies, which we will discuss next in
  \sref{sec:degen}.

\subsection{Degeneracies in Neutrino Observables}
\label{sec:degen}

There are a number of degeneracies and uncertainties that may
prevent  fully conclusive statements regarding the mapping from
detected signal to progenitor core structure.  These include, nuclear
EOS, rotation, viewing angle, distance, and neutrino oscillations
(including collective oscillations). We will discuss each one of these
consecutively and independently, although all may be relevant in a
generic situation.

\begin{figure}[t]
\centering
\includegraphics[width=0.97\columnwidth]{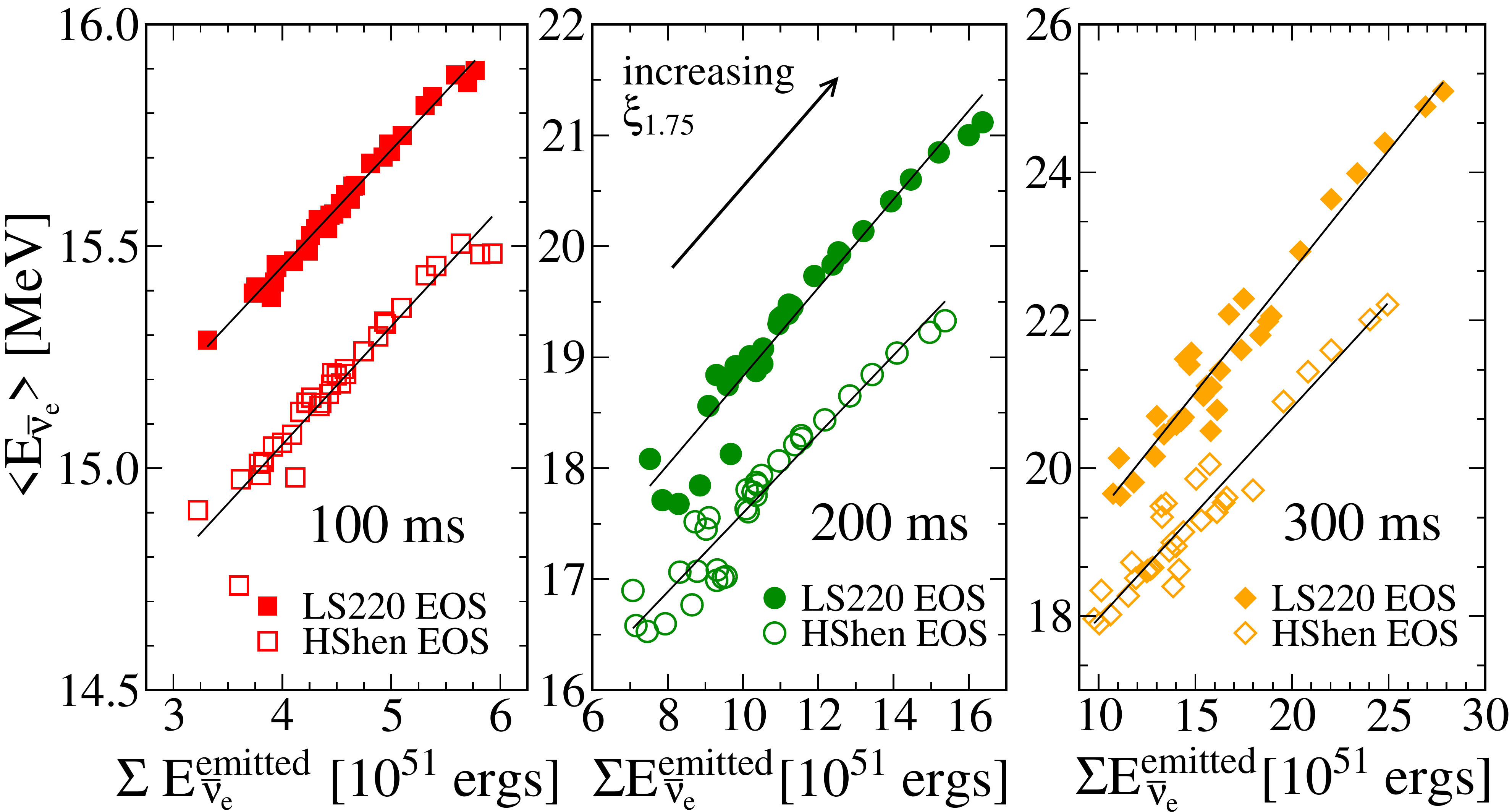}
\caption{Average $\bar{\nu}_e$ energy versus cumulative emitted
  $\bar{\nu}_e$ energy each EOS and
  progenitor.  This is shown for three postbounce times, 100\,ms (left
  panel); 200\,ms (center panel); and 300\,ms (right panel).  Solid
  (open) points correspond to the LS220 (HShen) EOS. The degeneracy
  between EOS and cumulative $\bar{\nu}_e$ energy is broken when the
  average energy is taken into account. We find that is a robust prediction
  across all progenitors and postbounce preexplosion
  times.}\label{fig:EOSdep}
\end{figure}

\emph{Nuclear EOS}: We have already briefly explored the dependence of
the neutrino observables on the nuclear EOS by taking two very
different EOS and comparing the emitted neutrino signal.  The total
number of interactions predicted to be detected from a given
progenitor model varies with EOS (cf., \fref{fig:cumul_nhits}) in such
a way that a high compactness model paired with the HShen EOS produces
the same number of interactions as a model with a slightly lower
compactness paired with a softer EOS, like the LS220. Hence, there is
a clear degeneracy between the progenitor compactness and the equation
of state. However, a water Cherenkov detector will not only detect a
given number of IBD interactions, but also their energy distribution.
From this distribution, and the detector response, one can work
backwards to reconstruct both the emitted average $\bar{\nu}_e$ energy
and the luminosity.  We show in \fref{fig:EOSdep} how knowledge of the
emitted energy spectrum can break the degeneracy with cumulative
emitted $\bar{\nu}_e$ energy and EOS, provided the distance to the
core collapse event is known (see the discussion of distance
uncertainties later in this section).  For each model and EOS, we plot
the average $\bar{\nu}_e$ energy as a function of cumulative emitted
$\bar{\nu}_e$ energy at three select postbounce times: 100, 200, and
300\,ms.  The data from each of the 32 progenitors fall on a unique
line that is parameterized by the compactness. In principle, provided
very large statistics and ultimate predictions of neutrino
observables, this allows one to break the EOS-progenitor degeneracy
between EOS.  In reality, achieving an instantaneous measurement of
the neutrino average energy to within a few \% will be difficult, and
there will likely be a large suite of nuclear EOS that could reproduce
the observables within error.

\begin{figure}[t]
\centering
\includegraphics[width=0.97\columnwidth]{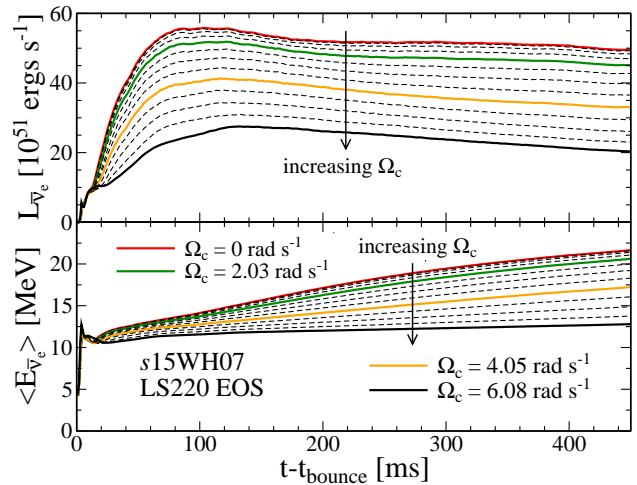}
\caption{Luminosities (top panel) and average energies (bottom panel)
  from 1.5D rotating core collapse simulations of model $s$15WH07 for
  various initial central angular velocities. We assign angular
  momentum to the $s$15WH07 progenitor via \eref{eq:rotationlaw} with
  values of $j_{16,\infty}$ ranging from 0 to 3 in 0.25 increments.
  This gives initial central rotation rates of 0 to $\sim$\,6.08 rad
  s$^{-1}$ in increments of $\sim$\,0.51 rad s$^{-1}$. In our
  simulations, increased rotation lowers both the neutrino luminosity
  and the average energy.}\label{fig:rotation}
\end{figure}

\emph{Rotation and Viewing Angle}: The effect of rotation on the
neutrino signal may be more difficult to disentangle from the effects
of progenitor structure. To explore the effect of rotation, we perform
1.5D simulations of rotating core collapse.  These simulations treat
rotation in a spherically symmetric way, using `shellular rotation'
\citep{thompson:05,oconnor:10}.  This approximation only captures the
spherically-averaged centrifugal effect of rotation on the matter. As
the amount of rotation is increased, the protoneutron star becomes
more and more centrifugally supported. Increased rotation leads to
lower densities and temperatures throughout the protoneutron
star. This in turn effects the neutrino signal. In
\fref{fig:rotation}, we plot the $\bar{\nu}_e$ luminosities and
average $\bar{\nu}_e$ energies determined from \code{nuGR1D} for 12
rotating core collapse simulations.  We use model $s$15WH07 paired
with the LS220 EOS.  The initial rotation rate is assigned via
\citep{oconnor:11}
\begin{equation}
j(r) =
j_{16,\infty}\left[1+\left(\frac{A_{M_\odot}}{r}\right)^2\right]^{-1}\times
10^{16}\ \mathrm{cm}^2\ \mathrm{s}^{-1}\,,\label{eq:rotationlaw}
\end{equation}
where $A_{M_\odot}$ is the radius that encloses 1\,$M_\odot$, which
for the $s$15WH07 progenitor is 703\,km, and $j_{16,\infty}$ is the
specific angular momentum at infinity, in units of
$10^{16}\ \mathrm{cm}^2\ \mathrm{s}^{-1}$. We vary $j_{16,\infty}$
from 0 to 3 in increments of 0.25. For the $s$15WH07 progenitor, the
initial central angular velocity is $\Omega_c = 2.03
j_{16,\infty}$\,rad\,s$^{-1}$, giving a range of $\Omega_c$ from 0 to
6.08\,rad\,s$^{-1}$. At bounce, the $j_{16,\infty}=1, \Omega_c =
2.03$\,rad\,s$^{-1}$ model has a rotation period of $\sim$\,2.4\,ms,
which decreases nearly linearly with increasing initial
$j_{16,\infty}$. Increased rotation leads to the overall reduction of
both the $\bar{\nu}_e$ luminosity and average energy. For example,
with $\Omega_c$ = 0, 2.03, 4.05, and 6.08\,rad\,s$^{-1}$ the
$\bar{\nu}_e$ luminosity at 100\,ms is $\sim$\,55, $\sim$\,52,
$\sim$\,40, and $\sim$\,25\,B\,s$^{-1}$, respectively. While we do not
show the $\nu_e$ and $\nu_x$ luminosities and average energies, they
follow the $\bar{\nu}_e$ trends. The only exception to this is the
$\nu_e$ neutronization burst where rotation in our 1.5D approach does
not significantly alter the $\nu_e$ luminosity. Our 1.5D approach to
rotation does not capture the anisotropy of the neutrino emission,
rather just the angle-averaged value. \cite{ott:08} considered a model
with an initial rotation rate of $\pi$\,rad\,s$^{-1}$, roughly
corresponding to our $j_{16,\infty}=1.5$ model. They found that the
ratio of polar to equatorial luminosity can be as large as 3--4 and
that average neutrino energies in polar regions can be harder by
$\gtrsim $1--2\,MeV than on the equator. We find that it is not
possible with the cumulative neutrino signals alone to break the
degeneracy between rotation and the progenitor star--EOS
combinations. The angle-averaged $\bar{\nu}_e$ luminosity and average
energy from a rotating system mimic the $\bar{\nu}_e$ luminosity and
average energy from a non-rotating system with a progenitor with lower
compactness and/or a stiffer EOS.  In addition, the rotational energy
stored in the neutron star remnant can lead to an underestimate of the
total emitted neutrino energy, again mimicking a lower-compactness
progenitor and/or a stiffer EOS. However, all hope is not lost:
stellar evolution and current constraints on pulsar birth spins (e.g.,
\citealt{heger:05, ott:06spin}) suggest that most massive stars will
be spinning too slowly for rotation to have a strong effect on
dynamics and neutrino signal. \cite{heger:05}, for example, predict a
presupernova rotation rate of 0.2\,rad\,s$^{-1}$ for a 15\,$M_\odot$
star, which is smaller than all rotation rates considered here.  If,
on the other hand, the progenitor is rapidly spinning, the degeneracy
in the neutrino signal may be broken by coincident observations in
gravitational waves that are able to constrain the rotation rate of
the collapse core \citep{ott:12a}.

\emph{Distance}: If the electromagnetic signal is blocked, by, for
example, the galactic center, we may not be able to obtain a reliable
distance to the supernova.  The $\bar{\nu}_e$ flux at Earth follows an
inverse square law, while the spectral distribution will not change
over the distances of relevance. This means that degeneracy with
distance is also hard to break with the IBD neutrino signal alone for
the following reason: A high-compactness progenitor at a large
distance can produce the same energy-integrated flux at Earth as a
low-compactness progenitor. Less neutrinos are emitted in the latter
but the flux has not been diluted as much due to the closer
distance. If the nuclear EOS is not known, the observed neutrino flux
and energy spectra could be associated with a range of nuclear
EOS--distance combinations. A progenitor at a large distance could
produce the same neutrino flux and energy distribution as a
lower-compactness progenitor with a softer EOS.  Here, the softer EOS,
which will give rise to an increased average energy,  
compensates for the decrease in the average energy produced by the
smaller compactness.

Future constraints on the nuclear EOS could break this degeneracy
since one could more reliably associate an observed energy
distribution with a particular progenitor compactness
(\fref{fig:lums_and_aveen}).  This may even provide a distance
estimate, as would a detection of the neutronization burst signal
\citep{kachelriess:05}.  It does not show a strong dependence on the
progenitor model, but is difficult to detect since the $\nu_e$ cross
sections in water Cherenkov detectors are much lower than the IBD
cross section, resulting in few interactions. Some detector materials
have significantly larger $\nu_e$ cross sections and would be better
suited to detect the neutronization burst, such as liquid argon
\citep{scholberg:12} and lead \citep{duba:08halo}.

\emph{Neutrino Oscillations:} First, considering only matter-induced
neutrino oscillations, in the normal mass hierarchy, the electron
antineutrino signal at Earth is a composite spectrum of
$\cos^2{(\theta_{12})}\sim$\,70\% of the original $\bar{\nu}_e$
neutrinos and $\sin^2{(\theta_{12})}\,\sim$\,30\% of the original
$\bar{\nu}_x$ spectrum, where $\theta_{12}$ is the mixing angle
between the mass eigenstates $1$ and $2$. In the inverted neutrino mass
hierarchy, the entire $\bar{\nu}_e$ signal is replaced with the
original $\bar{\nu}_x=\nu_x$ signal \citep{dighe:00}. In either case,
we still expect the total number of IBD interactions to increase with
the compactness parameter of the progenitor since both the
$\bar{\nu}_e$ and the $\nu_x$ luminosity increase.  If the hierarchy
of the neutrino mass eigenstates is not determined by neutrino
experiments before the next nearby core-collapse supernova, the early
postbounce, preexplosion neutrino luminosities may provide an answer
\citep{kachelriess:05,serpico:12}. This relies on the systematically
different rise times of the $\nu_x$ and $\bar{\nu}_e$ signals. We also
point out a degeneracy in disentangling the spectral properties of the
emitted neutrino spectra from the matter-induced oscillated neutrino
spectra observed in Earth detectors \citep{minataka:08}.

Much more troublesome are the collective neutrino oscillations that
arise from coherent neutrino-neutrino forward scattering.  Collective
oscillations are very sensitive to the energy spectra (both the
distribution and magnitude) of all neutrino flavors and the background
matter density. Since the governing equations are highly non-linear,
there are currently no simple analytic expressions predicting the
neutrino signal at Earth based on the output of core-collapse
simulations. Recent studies suggest that during the early postbounce,
preexplosion phase, collective neutrino oscillations may be suppressed
\citep{chakraborty:11prl, Chakraborty:2011gd, sarikas:12},
however the community has not yet reached consensus, see, e.g.,
\cite{cherry:12} and \cite{dasgupta:12a}.

\section{Discussion}
\label{sec:discussion}

The next nearby core-collapse supernova will be extremely well
observed in neutrinos. Super-Kamiokande alone will observe
$\sim$$7000$ electron antineutrinos from a typical core-collapse
supernova at a fiducial galactic distance of
$10\,\mathrm{kpc}$. Future detectors of the scale of the proposed
Hyper-Kamiokande may see in excess of $10^5$ interactions.  Such
high-statistics observations will provide rich information on the
neutrino signal. Comparison with theoretical model predictions will
allow to falsify or constrain a broad range of hypotheses in
core-collapse supernova astrophysics and nuclear/neutrino physics.
Unexpected signal features may lead to the discovery of new physics.

In this study, our focus has been on the imprint of the progenitor
star's structure on the neutrino signal in the postbounce preexplosion
phase of core-collapse supernovae. We have carried out a large set of
spherically-symmetric radiation-hydrodynamics simulations of core
collapse and the early postbounce phase with the goal
of studying trends in the neutrino signal with variations in
progenitor structure.

Our results show, in agreement with previous work (e.g.,
\citealt{burrows:83nat, liebendoerfer:01c, liebendoerfer:02b,
  liebendoerfer:03, liebendoerfer:04, thompson:03, kachelriess:05,
  buras:06b, serpico:12}), that the $\nu_e$ signal from the
neutronization burst emerging shortly after bounce has very little
progenitor dependence, due to the universal nature of homologous inner
core collapse.

The neutrino signal in the postbounce preexplosion phase is determined
primarily by the accretion luminosity of outer iron core, silicon
shell, and oxygen shell material. The postbounce accretion rate
depends on the inner structure of the progenitor at the presupernova
stage. Our results show that the preexplosion neutrino signal has an
essentially monotonic dependence on progenitor structure described by
a single parameter, the compactness $\xi_M \propto M / R(M)$ (where
$M$ is a typical baryonic mass reaching the center over the timescale
of interest). The greater a progenitor's $\xi_M$, the higher are the
emitted luminosities and average energies of all neutrino species.
Scaling in the same way is the total emitted energy in neutrinos over
the entire protoneutron star cooling phase for a given explosion time
at which accretion is shut off.  These trends are robust and
independent of the nuclear EOS. They are also rather insensitive to
the particular choice of the reference mass $M$ in $\xi_M$ as long as
it is in the range of typical neutron star baryonic masses ($\sim$$1.4
- 2.5\,M_\odot$) and we find $\xi_{1.75}$ to be a good choice.

\begin{figure}[t]
\centering
\includegraphics[width=0.97\columnwidth]{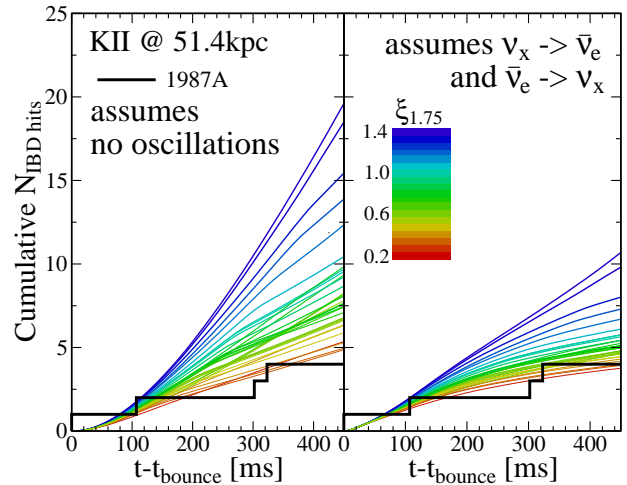}
\caption{Predictions of the cumulative IBD interactions in
  Kamiokande--II for a core-collapse event at 51.4\,kpc.  The left
  panel assumes no neutrino oscillations, while the right panel
  assumes a swapping of the $\bar{\nu}_e$ and $\nu_x$ spectra. The
  color coded lines denote different progenitors and are generated
  with \code{SNOwGLoBES} based on our simulations with the LS220
  EOS. More details are provided in the text.  We overlay the
  cumulative detected interactions observed in Kamiokande--II from SN
  1987A, assuming the first interaction denotes the time of
  core-bounce. We caution the reader by noting that the very low
  number statistics make it hard to draw any conclusions
  regarding the progenitor of SN1987A.}\label{fig:1987A}
\end{figure}

The monotonic dependence of the preexplosion neutrino emission on
progenitor compactness translates directly to the neutrino signal
observed by detectors, provided collective neutrino oscillations do
not lead to complicated swaps of flavor spectra that brake the
dependence of the observed signal on progenitor structure. Neutrino
observations of the next nearby core collapse event thus may, in
principle, allow quantitative constraints on the inner structure of
the progenitor star. As an example with real neutrino data, we
consider the early postbounce neutrino signal observed from SN 1987A
by the Kamiokande--II experiment \citep{hirata:87}. Of the eleven
interactions that were observed, the first four occurred within
323\,ms of each other. All interactions observed by Kamiokande--II are
consistent with being IBD interactions \citep{hirata:87}. We assume
that the first interaction occurs at the onset of the postbounce
phase, although the actual bounce time is likely somewhat earlier. In
\fref{fig:1987A}, we plot the cumulative number of detected
interactions observed from SN 1987A in the first 500\,ms along with
the \code{SNOwGLoBES} prediction from our simulations with the LS220
EOS using a 2.14\,kT water Cherenkov detector at 51.4\,kpc.  We use
the efficiencies quoted in \cite{burrows:88} and the smearing matrices
from the \code{SNOwGLoBES} detector configuration
\code{wc100kt30prct}.  To show the full range of the possible effects
of MSW neutrino oscillations, we show the expected number of
interactions assuming no oscillations and assuming a complete switch
of the $\bar{\nu}_e$ and $\nu_x$ spectra (as would be the case in the
inverted mass hierarchy; the normal hierarchy is a combination of
these two signals). During the accretion phase the expected number of
detected interactions from an oscillated $\nu_x$ spectrum can be
significantly smaller than the electron-type neutrino luminosity due
to the smaller neutrino luminosity in the accretion phase. While the
quantitative results obviously depend on neutrino oscillation details,
the qualitative trend with $\xi_M$ is unbroken. Comparing our
predictions with the interactions observed from SN 1987A one notes
(but must keep the very small-number statistics in mind) that either
the explosion must have occurred early in the postbounce phase and/or
the progenitor must have had a relatively low $\xi_{1.75}$.  Stated
another way, that data strongly disfavors a late-time explosion in a
high compactness model.  Both of these statements are broadly
consistent with the previous work of \cite{bruenn:87} and
\cite{burrows:88}. It is also not inconsistent with the proposed ZAMS
mass of $\sim$$18-20\,M_\odot$ for Sanduleak~$-69^\circ$~$202$ (e.g.,
\citealt{whw:02}), the blue supergiant progenitor star of SN
1987A. However, the mapping between ZAMS mass and stellar structure
(i.e., compactness) at the presupernova stage appears to be highly
non-monotonic (cf.~Fig.~\ref{fig:xis};
\citealt{whw:02,woosley:07}). This makes it very difficult to link the
observed neutrino signal to ZAMS mass without additional constraints
from classical astronomical observations in the electromagnetic
spectrum.

In the discussion of SN 1987A, we have ignored uncertainties regarding
nuclear EOS, rotation of the progenitor core, distance, and collective
neutrino oscillations. All affect the preexplosion neutrino signature
in ways that may be degenerate with variations in the compactness
parameter. However, as we have shown, large variations in the
stiffness of the nuclear EOS in the density and temperature regime
relevant in the preexplosion phase can be disentangled from $\xi_M$
via observations of the emitted neutrino spectrum. Rapid rotation,
which decreases both (angle-averaged) luminosities and average
neutrino energies, can be constrained by gravitational wave
observations of a galactic event (e.g., \citealt{ott:12a}).  Distance
uncertainties from electromagnetic observations, which translate into
uncertainties in the absolute luminosities, can be reduced by
exploiting the generic neutronization burst as a standard candle
\citep{kachelriess:05}. Collective neutrino oscillations are not yet
fully understood and have not been directly incorporated in neutrino
radiation-hydrodynamics simulations. They may lead to one or multiple
energy-dependent swaps of spectra between flavors and could thus
complicate the mapping between neutrino signal and compactness
parameter. However, the recent understanding suggests that collective
oscillations may not be significant in the preexplosion phase
(\citealt{chakraborty:11prl, Chakraborty:2011gd, sarikas:12};
but: \citealt{cherry:12} and \citealt{dasgupta:12a}).

Our goal with this study was to highlight overall trends of the
preexplosion neutrino signal with progenitor structure in the limit of
spherical symmetry. We expect these overall trends to be robust and
to carry over to the multi-dimensional case.  We did not aim at making
precise and robust quantitative predictions for any individual
model. These are not possible with our current spherically-symmetric
radiation-hydrodynamics treatment, which neglects inelastic
scattering, redshift and velocity dependent terms (e.g.,
\citealt{lentz:12b,lentz:12a}). Rotation is included only in an
angle-averaged manner and other important multi-dimensional dynamics,
in particular convection and the standing-accretion shock instability
\citep{buras:06b,ott:08,marek:09,marek:09b,lund:10,brandt:11,ott:12a},
cannot be captured.  Future work to remove these limitations will be
necessary to produce the reliable signal predictions necessary for
drawing quantitative conclusions from neutrino observations of the
next nearby core-collapse supernova. However, even with fully
realistic and complete simulation codes, large parameter studies will
be necessary to account for prevailing uncertainties in the nuclear
EOS and/or neutrino interaction physics.

\section*{Acknowledgements}

We acknowledge helpful discussions with and input from John Beacom,
Adam Burrows, Luc Dessart, Ken Nomoto, Ryan Patterson, Kate Scholberg,
Mark Vagins, and Stan Woosley.  CDO thanks the Kavli Institute for the
Physics and Mathematics of the Universe for hospitality while work on
a draft of this article was carried out. The computations were
performed at Caltech's Center for Advanced Computing Research on the
cluster ``Zwicky'' funded through NSF grant no.\ PHY-0960291 and the
Sherman Fairchild Foundation. Furthermore, computations were performed
on Louisiana Optical Network Infrastructure computer systems under
allocation loni\_numrel06, on the NSF XSEDE Network under allocation
TG-PHY100033, and on resources of the National Energy Research
Scientific Computing Center, which is supported by the Office of
Science of the U.S. Department of Energy under Contract
No. DE-AC02-05CH11231.  EOC is supported in part by a post-graduate
fellowship from the Natural Sciences and Engineering Research Council
of Canada (NSERC). This research is supported in part by the National
Science Foundation under grant nos. AST-0855535 and OCI-0905046, by
the Alfred P. Sloan Foundation, and by the Sherman Fairchild
Foundation.

\end{document}